\documentclass[useAMS,usenatbib]{mn2e}

\usepackage{txfonts}
\usepackage{graphicx}
\usepackage{times}
\usepackage{natbib}
\usepackage{subfigure}
\usepackage{footnote}
\usepackage{lscape}
\usepackage{multirow}
\usepackage{longtable}
\usepackage{rotating}
\usepackage{float}

\title[VLT survey of Wolf-Rayet stars in NGC 7793]{A Very Large Telescope imaging and spectroscopic
survey of the Wolf-Rayet population in NGC 7793 \thanks{Based on observations made with ESO telescopes at the 
Paranal observatory under program ID 081.B-0289 and 
archival NASA/ESA Hubble Space Telescope datasets, 
obtained from the ESO/ST-ECF Science Archive Facility.}}
\author[J. L. Bibby  and P. A. Crowther]{J. L. Bibby \thanks{Email: j.bibby@sheffield.ac.uk} \& P. A. Crowther \\
University of Sheffield, Department of Physics \&
              Astronomy, Hicks Building, Hounsfield Rd, Sheffield, S3
              7RH} 

\begin{document}

\date{}

\pagerange{\pageref{firstpage}--\pageref{lastpage}} \pubyear{2002}

\maketitle

\label{firstpage}

\begin{abstract}
We present a VLT/FORS1 imaging and spectroscopic survey of the
Wolf-Rayet (WR) population in the Sculptor group spiral galaxy NGC
7793. We identify 74 emission line candidates from archival narrow-band
imaging, from which 39 were observed with the Multi Object
Spectroscopy (MOS) mode of FORS1. 85\% of these sources displayed WR
features. Additional slits were used to observe H\,{\sc ii} regions,
enabling an estimate of the metallicity gradient of NGC 7793 using
strong line calibrations, from which a central oxygen content of
$\log$ (O/H) + 12 = 8.6 was obtained, falling to 8.25 at R$_{\rm
  25}$. We have estimated WR populations using a calibration of line
luminosities of Large Magellanic Cloud stars, revealing $\sim$27 WN
and $\sim$25 WC stars from 29 sources spectroscopically
observed. Photometric properties of the remaining candidates suggest
an additional $\sim$27 WN and $\sim$8 WC stars. A comparison with the
WR census of the LMC suggests that our imaging survey has identified
$\sim$80\% of WN stars and $\sim$90\% for the WC subclass. Allowing
for incompleteness, NGC 7793 hosts $\sim$105 WR stars for which
N(WC)/N(WN)$\sim$0.5.
From our spectroscopy of H\,{\sc ii} regions in NGC~7793, we revise
the global H$\alpha$ star formation rate of Kennicutt et al. upward by
50\% to 0.45 M$_{\odot}$ yr$^{-1}$. This allows us to obtain
N(WR)/N(O) $\sim$0.018, which is somewhat lower than that resulting
from the WR census by Schild et al. of another Sculptor group spiral
NGC 300, whose global physical properties are similar to NGC 7793.
Finally, we also report the fortuitous detection of a bright ($m_{\rm
  V}$ = 20.8 mag) background quasar Q2358-32 at $z \sim 2.02$
resulting from C\,{\sc iv} $\lambda$1548-51 redshifted to the
$\lambda$4684 passband.
\end{abstract}


\section{Introduction}

Classical Wolf-Rayet (WR) stars are helium burning stars descended
from massive O stars. Their strong stellar winds produce a unique
broad emission-line spectrum, making WR stars easily identifiable in
both Local Group \citep{Massey&Johnson1998} and more distant
star-forming galaxies \citep{Conti&Vacca1990}. WR stars contribute
significantly to the chemical evolution of the interstellar medium
(ISM) via stellar winds and core-collapse supernova
(ccSNe, \citealt{Dray&Tout2003}). Indeed WR stars are believed to be the
progenitors of Type Ib/c supernova and some long Gamma-Ray Bursts
(GRBs), however a direct observational link is yet to be established
\citep{Woosley&Bloom2006}.

Wolf-Rayet stars can be divided into subtypes that are
nitrogen-rich (WN) or carbon-rich (WC). Metal-rich environments are
observed to favour WC stars due to stronger, metal-driven winds during
both the WR phase \citep{Crowther2002} and the progenitor O star phase
\citep{Mokiem2007}, while we expect to find a higher fraction of WN
stars in metal-poor environments \citep{Massey&Johnson1998}. We can
investigate the distribution of WR subtypes with respect to
metallicity by studying galaxies spanning a range of
metallicities. Indeed, many spiral galaxies possess a super-solar nuclei and
sub-solar outer regions (eg. \citealt{Pagel&Edmunds1981,
  Magrini2007}).

It is thought that WN and WC stars are the progenitors of Type Ib and
Ic SNe, respectively. The advent of 8-m class telescopes has allowed
searches for WR populations to move beyond the Local Group
\citep{Schild2003}. The identification of a Type Ib/c supernova
progenitor is the long-term aim of our survey. The survey consists of
10 nearby star-forming galaxies, and one dwarf irregular galaxy, which
were largely chosen based on criteria such as distance, star-formation
rate and orientation. To date five galaxies in our sample have been
completed (\citealt{Schild2003, Hadfield2005, Hadfield2007, 
  CrowtherBibby2009} \& this work), whilst three are underway, and
three are in the preliminary stages.

By surveying $\sim$10 galaxies within 10\,Mpc, our overall aim is to
produce a complete catalogue of $\geq$10$^{4}$ WR stars which can be
referred to when a Type Ib/c supernova occurs. O stars have lifetimes
of 3--10\,Myr, of which $\sim$0.5\,Myr is spent in the WR phase
\citep{Crowther2007}. Given this short lifetime, statistically we
would expect at least one of the stars in our sample to undergo
core-collapse producing a Type Ib (H-poor) or Type Ic (H, He-poor) SNe
within the next few decades. \citet{Kelly2008} investigate the
location of different classes of supernovae relative to the light
distribution of the host galaxy which supports different progenitors
for Type Ib and Ic SNe. \citet{LeLoudas2010} extend this investigation
to the distribution of WR subtypes with respect to the light
distribution of two galaxies (M83 and NGC 1313) in our sample. They
find WC stars to favour the brighter regions, consistent with the
prediction that WC stars are progenitors of Type Ic SNe. Moreover,
early-type WN (WNE) stars are found to be more consistent with the
distribution of Type Ib SNe, and are ruled out as Type Ic SNe
progenitors.

NGC 7793 is a SA(s)d galaxy \citep{deVaucouleurs1991} that is part of
the Sculptor group of galaxies at a distance of 3.91\,Mpc
\citep{Karachentsev2003}. Despite its relatively low star-formation
rate (0.3\,M$_{\odot}$yr$^{-1}$, \citealt{Kennicutt&Lee2008}) its low
distance and favourable orientation make it an appropriate addition to
our galaxy survey. Previous spectroscopic observations (using the
Anglo-Australian 4m telescope) of 4 H\,{\sc ii} regions within
NGC~7793 have detected weak, broad He\,{\sc ii}\,$\lambda$\,4686 emission
\citep{Chun1983}. However no comprehensive WR survey has been
undertaken to date. Previous, albeit few, observations of H\,{\sc ii}
regions within NGC~7793 suggest that it has a shallow metallicity
gradient \citep{Webster1983}.

In this paper we use Very Large Telescope (VLT) optical imaging and
spectroscopy, combined with archival VLT and Hubble Space Telescope
(HST) images to determine the massive stellar content of
NGC~7793. Details of observations of NGC~7793 are presented in Section
2, including details of WR candidate selection. In Section 3 we
discuss the properties of the nebular, whilst stellar properties and
WR subtypes are determined in Section 4. Section 5 provides a
comparison between ground and space-based observations and addresses
the completeness of our survey in relation to WR stars in the Large
Magellanic Cloud (LMC). A discussion of Giant H\,{\sc ii} regions
follows in Section 6, whilst Section 7 reports the serendipitous
detection of a background quasar Q2358-32. Section 8 discusses the
global WR population of NGC~7793 and is compared with the WR content
of NGC 300, another Sculptor group spiral, and other nearby
galaxies. The paper concludes with a brief summary in Section 9.

\section{Observations \& Data Reduction}

Imaging and spectroscopy of NGC 7793 were obtained in July and
September 2008 using the Focal Reduced/Low Dispersion Spectrograph \#1
(FORS1) mounted at the European Southern Observatory (ESO) VLT. In
addition, archival VLT/FORS1 images were used and supplemented with
Hubble Space Telescope (HST) Advanced Camera for Surveys (ACS) optical
images.

\begin{table}
\caption{VLT/FORS1 observational log for NGC 7793. The seeing is
  calculated from the FWHM of bright, unsaturated, stars in the field
  of view.}
\begin{tabular}{c@{\hspace{1mm}}c@{\hspace{1mm}}c@{\hspace{1.5mm}}c@{\hspace{1mm}}c@{\hspace{1mm}}c@{\hspace{1mm}}}
\hline
Date & Filter/  & $\lambda_{c}$ & Proposal ID/ & Exposure & Seeing \\
     & mask ID  &     ($\AA$)  & PI           & time (s)  & (arcsec)    \\ 
\hline
\multicolumn{5}{c}{Imaging}\\
\hline
2002-11-01  & He\,{\sc ii} & 4684 & 70.D-0137(A) & 800 & 1.3  \\
 \vspace{1mm}      & [O\,{\sc iii}]/6000 & 5100 &/ Royer & 800 & 1.1  \\

2008-07-05         & H $\alpha$ & 6563 &   081.B-0289(D)& 270 & 0.75 \\
                   & H $\alpha$/4500 & 6665 &   / Crowther & 270 & 0.75  \\
                   & B high       & 4440 &              & 270 & 0.75 \\
                   & V high       & 5570 &              & 270 & 0.75 \\
\hline
\multicolumn{5}{c}{MOS Spectroscopy} \\
\hline
2008-07-09     & MASK 1       & 5850 & 081.B-0289(E) & 1650 & 1.25 \\
               & MASK 2       & 5850 &/ Crowther     & 1650 & 1.25 \\
2008-09-23     & MASK 3       & 5850 &               & 1650 & 1.25 \\
\hline

\end{tabular}
\label{observation log}
\end{table}

\begin{figure}
   \centering
   \includegraphics[width=0.9\columnwidth]{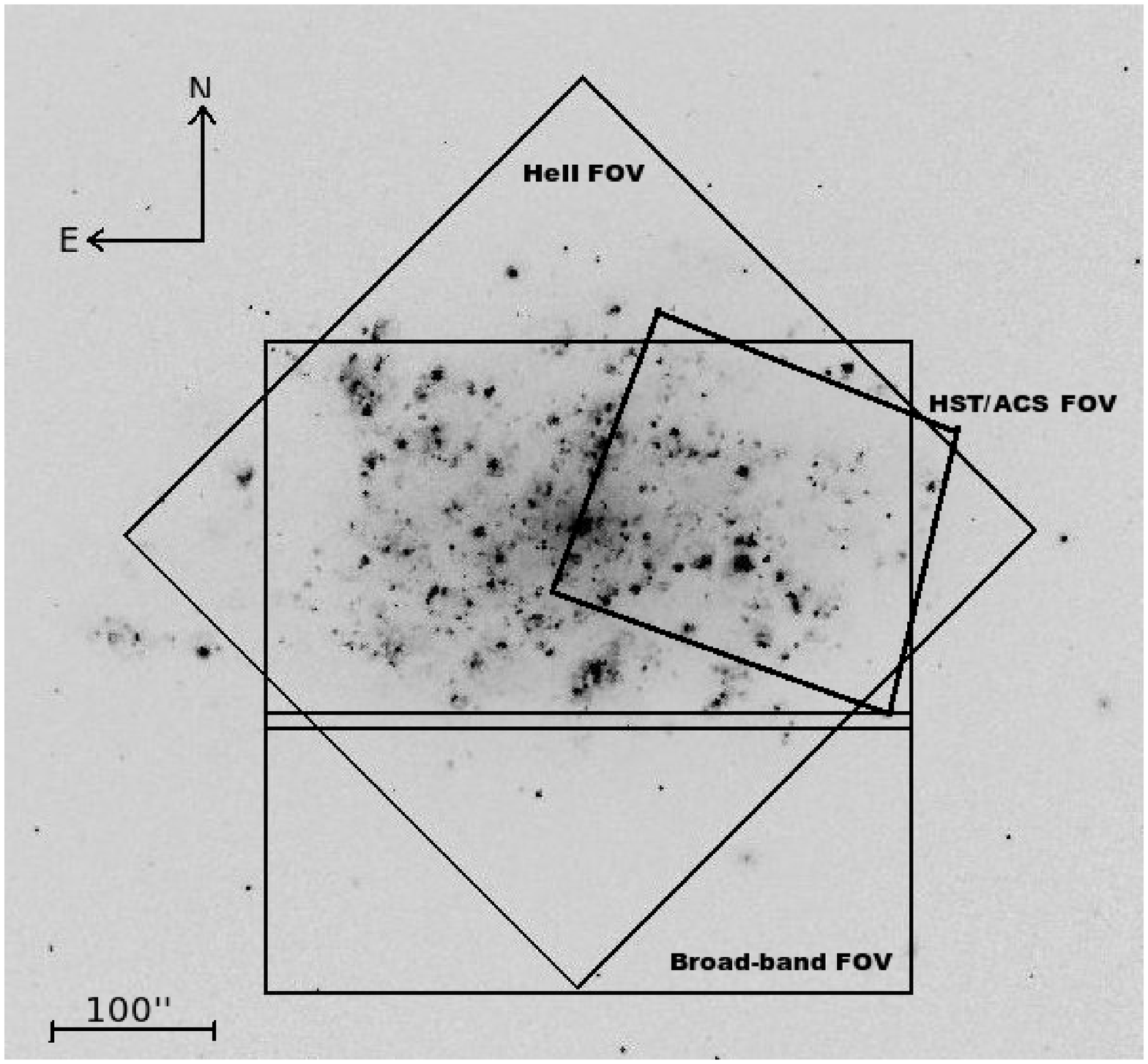}
     \caption{CTIO wide-field H$\alpha$ image of NGC~7793
       \citep{Kennicutt&Lee2008}, showing the field of view (FOV) of
       the B,V, H$\alpha$ images (``broad band''), archival He\,{\sc
         ii} \& [O\,{\sc iii}] images, and the HST/ACS F555W images.}
                   \label{fov}
   \end{figure}

\subsection{VLT/FORS1 Imaging}
Broad-band B- and V-high throughput images plus narrow-band H$\alpha$
and continuum images were obtained in July 2008 (270 seconds exposure)
under good conditions (FWHM = 0.75 $\arcsec$) using the two E2V
blue-sensitive chips on FORS1 with a 6.8 $\times$ 6.8 arcminute field
of view and a pixel scale of 0.25 arcsec pixel$^{-1}$. A log of the
observations is shown in Table \ref{observation log}. These images
were reduced using standard procedures (bias subtraction \& flat
fielding) within \textsc{IRAF} \citep{Tody1986}.

In addition, archival narrow-band He\,{\sc ii}\,$\lambda_{c}$=4684$\AA$
imaging of NGC~7793 were available via the ESO archive along with
[O\,{\sc iii}] continuum ($\lambda_{c}$=5100$\AA$) images, offset from the
[O\,{\sc iii}]$\lambda$5007 line by 6000 km/s. These were taken with
VLT/FORS1 (program ID 70.D--0137(A)) under average seeing conditions
(1.1--1.3 $\arcsec$). The 6.8 $\times$ 6.8 arcminutes (0.2 arcsec
pixel$^{-1}$) field of view was covered by one 2k $\times$ 2k Tektronix
CCD. Due to the different detectors used on FORS1 the archival images
have a different field of view to our own images, as shown in Figure
\ref{fov}.

\subsection{HST/ACS Imaging}
Ground-based data were supplemented with high resolution archival
HST/ACS imaging taken from program 9774 (PI: Larsen,
\citealt{Mora2009}). These images were obtained with the Wide Field
Channel (WFC) using the F555W filter with an exposure time of 680
seconds. The much improved spatial resolution allows us to accurately
locate the source of the He\,{\sc ii} emission identified from FORS1 imaging
in some instances (see Section 5.2). Other regions of NGC~7793 are
covered by numerous HST pointings, however V-band images are only
available using the F555W filter of the region highlighted in Figure
\ref{fov}.

\subsection{FORS1 Photometry \& Zero Points}

Aperture photometry was performed using the \textsc{daophot} routine
within \textsc{iraf}. The photometric errors for objects within the
narrow-band image range from $\pm$0.03 mag for bright
(m$_{4684}$\,=\,20 mag) sources to $\pm$0.34 mag for the faintest
(m$_{4684}$\,=\,25 mag) sources. For the broad-band images typical
photometric errors range from $\pm$0.04 to 0.25 mag for bright
(m$_v$\,=\,20\,mag) and faint (m$_v$\,=\,25\,mag) sources
respectively. Zero points for the broad-band images were determined by
comparing aperture photometry, obtained using the \textsc{starlink}
package \textsc{gaia} \citep{Draper2009}, of standard stars within the
TPhe and PG1323-086 fields \citep{Stetson2000}. The zero point for the
narrow-band H$\alpha$ image was obtained in a similar way using the
spectrophotometric standard star LTT~377 \citep{Hamuy1992,
  Hamuy1994}. The systematic error on the zero point is $\pm$0.15 mag.

The calibration of the archival $\lambda$\,4684 and $\lambda$\,5100
images proved more challenging as standard stars were not obtained
together with the NGC~7793 images. Given the central wavelength of the
$\lambda$\,5100 filter we assumed that the $\lambda$\,5100 magnitude
corresponds to the average of the B and V magnitudes from which a zero
point was obtained. To calculate the zero point for the
$\lambda$\,4684 images we adopted m$_{4684}$-m$_{5100}$=0\,mag on
average. The formal error on the overall zero point applied to the
$\lambda$\,5100 and $\lambda$\,4684 images was $\pm$0.18 mag.

With the exposure times shown in Table \ref{observation log} the
archival VLT/FORS1 He\,{\sc ii}\,$\lambda$~4684 imaging are complete
to m$_{4684}$\,=\,22.85 mag corresponding to
M$_{4684}$\,=\,--5.76$\pm$0.18 mag for a distance of 3.91\,Mpc
\citep{Karachentsev2003} and an average extinction of
E(B-V)\,=\,0.179$\pm$0.024 mag (see Section 3.1). Similar values were
obtained for the $\lambda_{c}$\,=\,5100 continuum images.

\subsection{Source Selection}

From our archival narrow-band imaging we identified 74 emission line
sources using the ``blinking'' method pioneered by \citet{Moffat1983}
and \citet{Massey1983}.  \textsc{daophot} photometry revealed that
$\sim$25\% of our sources had multiple components. All sources and
sub-regions are listed in Table \ref{sources} together with the
corresponding photometry, we label the brightest components `a',
followed by `b' etc. In 63 of the WR candidate regions, at least one
of the multiple components showed an excess at $\lambda$\,4684
(He\,{\sc ii}/C\,{\sc iii} emission) with respect to the
$\lambda$\,5100 continuum emission up to 2.43 mag. An additional 10
regions could only be detected in the $\lambda_{c}$\,4684
filter. \textsc{daophot} could not achieve photometry of one candidate
(\#14) identified from the ``blinking'' method due to the spatially
extended nature of the $\lambda_{c}$\,4684 emission.

\begin{figure}
   \centering
   \includegraphics[width=0.9\columnwidth]{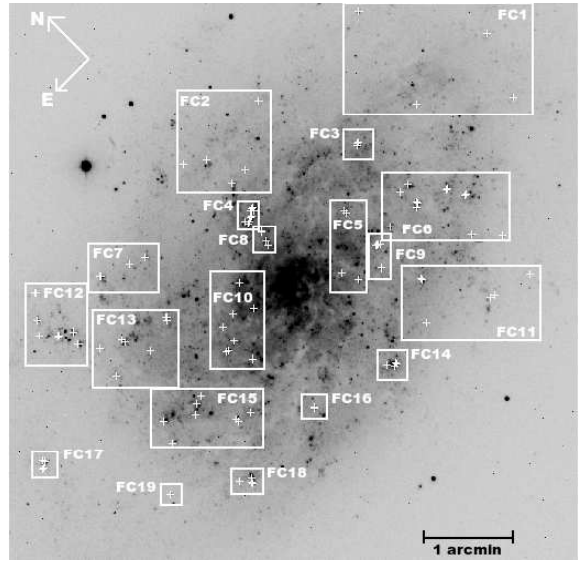}
     \caption{6.8 $\times$ 6.8 arcminute VLT/FORS1 He\,{\sc ii}
       finding chart showing the location of WR candidates. More
       detailed, individual finding charts can be found online.}
                   \label{wrlocation}
   \end{figure}

Higher spatial resolution broad-band VLT imaging, obtained at the same
time as the spectroscopy, enabled B-V colours to be determined. A
colour-cut of E(B-V)$\leq$0.4 mag was applied for all sources that lay
within the field of view of the broad-band image (recall Figure
\ref{fov}) to exclude very red sources. The spatial location of the WR
candidates in NGC~7793 is shown in Figure \ref{wrlocation}, with the
corresponding finding chart ID listed in Table~\ref{sources}. More
detailed finding charts are available electronically (Appendix B). The
deprojected distance from the centre of the galaxy, r/R$_{25}$, was
calculated using an inclination of $\textit{i}$=53.7$^{\circ}$ and
position angle of the major axis PA=99.3$^{\circ}$
\citep{Carignan1990}, with R$_{25}$=4.65 arcmin
\citep{deVaucouleurs1991, Corwin1994}.

\begin{table*}
\caption[]{ Catalogue of WR  candidates in NGC 7793 ordered by Right Ascension. 
Absolute   magnitudes are derived using a distance of 3.91 Mpc
  \citep{Karachentsev2003}. Where spectra has been obtained the
  derived E(B-V) is used. For those sources with no nebular lines seen
  an average E(B-V)=0.18 mag is used.  For cases without spectroscopic observation, we
provide photometric classifications (in parentheses). Association with 
H\,{\sc ii} regions relates to the catalogue of \citep{Davoust1980}.}  
\label{sources}
\begin{tabular}{
l@{\hspace{2mm}}
c@{\hspace{2mm}}
c@{\hspace{2mm}}
c@{\hspace{2mm}}
r@{\hspace{1mm}}
l@{\hspace{2mm}}
r@{\hspace{1mm}}
l@{\hspace{2mm}}
r@{\hspace{1mm}}
l@{\hspace{2mm}}
r@{\hspace{1mm}}
l@{\hspace{2mm}}
c@{\hspace{2mm}}
c@{\hspace{2mm}}
c@{\hspace{2mm}}
c@{\hspace{2mm}}
c@{\hspace{2mm}}
c}
\hline\noalign{\smallskip}
  ID & RA       &   Dec    & r/$R_{25}$ & \multicolumn{2}{c}{m$_{V}$} & \multicolumn{2}{c}{m$_B$--m$_{V}$} & \multicolumn{2}{c}{m$_{4684}$}  & 
\multicolumn{2}{c}{m$_{4684}$--m$_{5100}$}  & E(B-V)  &  
M$_V$ & M$_{4684}$  &  Spectral      & H\,{\sc ii} & Finding \\ 
     & J2000    & J2000    & & mag     & $\pm$ & mag        & $\pm$ & mag     & $\pm$ & mag &$\pm$  & mag & mag & mag&  Type         &  region    & Chart \\
\noalign{\smallskip}\hline
1   & 23:57:31.543 &  -32:35:00.29 &0.88& n/a     &    & n/a   &     & 20.28 &0.02& --0.11 &0.07   & 0.18 & n/a       & --8.33  & HII region &  D1 & 1 \\
2   & 23:57:33.067 &  -32:35:45.77 &0.76& 20.96   &0.07& --0.15& 0.08& 20.69 &0.03& --0.13 &0.07   & 0.18 & --7.56     & --7.92  & (not WR?) & -- & 1 \\
3   & 23:57:35.837 &  -32:33:43.87 &0.95& 20.40   &0.09& --0.01& 0.09& 20.12 &0.05& --0.28 &0.06   & 0.18 & --8.12     & --8.49  & HII region & D2 & 1 \\
4   & 23:57:37.240 &  -32:35:00.03 &0.62& 22.48   &0.07& --0.12&0.08 & 21.19 &0.02& --1.40 &0.08   & 0.13  & --5.88     & --7.24 & WC4--6 &  -- & 1 \\
5b  & 23:57:38.858 &  -32:36:11.06 &0.53& 20.41   &0.07& --0.25&0.12 & 20.15 &0.03& --0.01 &0.07   & 0.18  & --8.11    & --8.45 & (not WR?) & D8 & 6 \\
\smallskip
5a  & 23:57:38.942 &  -32:36:11.27 &0.53& 19.53   &0.08&   0.14&0.10 & 19.45 &0.03& --0.02 &0.04   & 0.18  & --8.99    & --9.16 & (not WR?) & D8 & 6 \\
6   & 23:57:39.049 &  -32:36:50.94 &0.65& 20.11   &0.06&   0.43&0.10 & 20.23 &0.03& --0.06 &0.07   & 0.18  & --8.41    & --8.38 & HII region & -- & 6\\
7a  & 23:57:39.381 &  -32:35:58.98 &0.49& 20.25   &0.06& --0.28&0.11 & 20.00 &0.03& --0.15 &0.07   & 0.18  & --8.27    & --8.61 & HII region & D10 & 6 \\
7b  & 23:57:39.450 &  -32:35:59.03 &0.48& 20.81   &0.06& --0.16&0.17 & 21.01 &0.06& --0.23 &0.18   & 0.18  & --7.71    & --7.60 & HII region & D10 & 6 \\
8   & 23:57:39.532 &  -32:37:24.42 &0.79& 20.97   &0.01& --0.21&0.21 & 20.72 &0.01& --0.43 &0.03   & 0.18  & --7.55    & --7.89 & WN10 & -- & 11 \\ 
\smallskip
9   & 23:57:40.257 &  -32:36:35.08 &0.55& 21.12   &0.01& --0.22&0.03 & 20.29 &0.01& --0.79 &0.14   & 0.00  & --6.84    & --7.67 & WN8 & D11& 6 \\
10  & 23:57:40.827 &  -32:35:36.48 &0.40& 20.17   &0.07&   0.03&0.13 & 19.29 &0.05& --0.65 &0.30   & 0.13$^{3}$  & --8.18    & --9.14 & WN2--4b  & D12 & 6 \\
11c & 23:57:41.105 &  -32:34:50.26 &0.47& 20.73   &0.07&   0.06&0.11 & 20.21 &0.05& --0.35 &0.09   & 0.00  &           & --7.75 & HII region & D14 & 3\\
11b & 23:57:41.197 &  -32:34:50.34 &0.47& 20.12   &0.10& --0.06&0.13 & 20.15 &0.08& --0.24 &0.09   & 0.00  & --7.84    & --7.81 & (WN?)     & D14 & 3 \\
11a & 23:57:41.250 &  -32:34:51.19 &0.46& 19.20   &0.08&   0.07&0.13 & 19.38 &0.04& --0.22 &0.06   & 0.00  & --8.76    & --8.58 & HII region & D14  & 3 \\
\smallskip
12b & 23:57:41.158 &  -32:35:50.17 &0.40& 20.17   &0.06&   0.11&0.11 & 20.01$^{1}$ &0.04$^{1}$&        &       & 0.18  & --8.35    & --8.60$^{1}$ & (not WR?)& D13 & 6 \\
12a & 23:57:41.165 &  -32:35:51.11 &0.40& 19.57   &0.09& --0.09&0.12 & 19.08 &0.05& --0.31 &0.11   & 0.18  & --8.95    & --9.53 & (WN?) & D13 & 6\\
13  & 23:57:41.340 &  -32:35:52.79 &0.39& 19.32   &0.05&   0.11&0.17 & 19.20 &0.03& --0.10 &0.03   & 0.18  & --9.20    & --9.41 & (Not WR?) & D13 & 6 \\
14  & 23:57:41.417 &  -32:35:35.97 &0.37& 21.27   &0.07&   0.05&0.07 & --    &    & --      &      & 0.30  & --7.61    &         & WC4  & D15 & 6 \\
15  & 23:57:41.738 &  -32:37:17.16 &0.71& 22.28   &0.02& --0.06&0.18 & 21.49 &0.02& --0.76 & 0.03  & 0.18  & --6.24    & --7.12 & (WN?) & D17 & 11 \\
\smallskip
16  & 23:57:42.040 &  -32:37:15.88 &0.70& 21.35   &0.06& --0.21&0.08 & 20.55 &0.04& --0.64 &0.04   & 0.26  & --7.43    & --8.36 & WN2--4 & D17 & 11 \\
17  & 23:57:43.254 &  -32:35:49.49 &0.31& 19.84   &0.08& --0.07&0.10 & 19.63 &0.04& --0.24 &0.04   & 0.18  & --8.68    & --8.98 & (WN?)& D20 & 6 \\
18  & 23:57:43.434 &  -32:33:38.17 &0.74& 22.05   &0.07& --0.01&0.07 & 21.54 &0.04& --0.37 &0.09   & 0.18  & --6.47    & --7.07 & (WN?) & -- & 2 \\
19b & 23:57:44.072 &  -32:36:31.77 &0.44& 20.38   &0.06& --0.27&0.08 & 19.68 &0.04& --0.43 &0.14   & 0.18  & --8.14    & --8.93 & (WN?) & D23 & 11 \\ 
19a & 23:57:44.092 &  -32:36:32.53 &0.44& 19.82   &0.03& --0.01&0.07 & 19.36 &0.04& --0.66 &0.05   & 0.18  & --8.70    & --9.25 & (WN?) & D23 & 11 \\ 
\smallskip
20  & 23:57:44.329 &  -32:35:53.13 &0.27& 22.50   &0.05& --0.04&0.18 & 20.98 &0.02& --1.51 &0.03   & 0.18  & --6.02    & --7.63 & WC4 & D25 & 9  \\ 
21a & 23:57:44.416 &  -32:35:17.09 &0.25& 19.64   &0.11&   0.35&0.15 & 19.70 &0.04& --0.17 &0.23   & 0.18  & --7.85    & --8.59 & (not WR?) & -- & 5 \\ 
21b & 23:57:44.416 &  -32:35:19.87 &0.24& 20.89   &0.04& --0.10&0.01 & 20.02 &0.07& --0.17 &0.18   & 0.18  & --7.63    & --8.11 & (not WR?) & --  & 5 \\ 
22b & 23:57:44.413 &  -32:35:52.01 &0.27& 21.51$^{2}$&0.07$^{2}$& &  & 21.16$^{1}$&0.07$^{1}$   &    &       & 0.18  & --7.19$^{2}$ & --7.44$^{1}$ & HII region & D25 & 9\\
22a & 23:57:44.491 &  -32:35:51.87 &0.26& 20.10   &0.06& --0.11&0.11 & 19.85 &0.04& --0.32 &0.10   & 0.18  & --8.42    & --8.76  & HII region & D25 & 9 \\
22c & 23:57:44.588 &  -32:35:51.56 &0.26& 21.43$^{2}$&0.10$^{2}$& &  & 21.26$^{1}$&0.05$^{1}$   &   &       & 0.18  & --7.27$^2$ & --7.35$^{1}$ & (WN?) & D25  & 9 \\
23  & 23:57:45.610 &  -32:36:56.14 &0.55& 23.15   &0.03& --0.19& 0.06& 21.07 & 0.02& --2.43 & 0.05 & 0.14  & --5.21    & --7.35 & WC4 & -- & 11 \\ 
24  & 23:57:46.647 &  -32:36:00.28 &0.24& 21.90   &0.02&   0.11& 0.03& 21.18 &0.02 & --0.77 & 0.05 & 0.18  & --6.62    & --7.43 & WC4 & D36 & 5  \\
25  & 23:57:46.805 &  -32:34:06.49 &0.50& 21.81   &0.04& --0.06& 0.05& 20.44 &0.03 & --1.09  &0.05 & 0.18  & --6.71    & --8.17 & WC4 & -- & 2\\
26  & 23:57:47.049 &  -32:35:48.11 &0.17& 22.49   &0.06& --0.16& 0.09& 21.44 &0.03&  --1.01  &0.10 & 0.17$^{3}$  & --5.99    & --7.12 & WC4--6 & --  & 5\\
\smallskip
27  & 23:57:47.868 &  -32:34:06.32 &0.49& 22.42   &0.08& --0.11&0.12 & 21.49 &0.04& --0.55   &0.07 & 0.18$^{3}$  & --6.10    & --7.12 & WC4 & D51 & 2\\
28  & 23:57:47.933 &  -32:33:41.44 &0.63& 21.22   &0.07& --0.10&0.54 & 20.74 &0.03& --0.28   &0.07 & 0.18  & --7.31    & --7.87 & (WN?) & D50 & 2 \\
29b & 23:57:48.064 &  -32:34:29.28 &0.35& 19.53   &0.05& --0.14& 0.30& 19.12 &0.03& --0.08   &0.19 & 0.18  & --8.99    & --9.42 & (not WR?) & D52 & 4 \\
29a & 23:57:48.184 &  -32:34:29.49 &0.34& 19.08   &0.01& 0.08  & 0.03& 18.85 &0.02& --0.04   &0.06 & 0.18  & --9.44    & --9.76 & (not WR?)& D52 & 4 \\
29c & 23:57:47.943 &  -32:34:29.45 &0.35& 20.43   &0.07& 0.09  & 0.13& 19.90 &0.05& --0.16   &0.09 & 0.18  & --8.09    & --8.71 & (not WR?)& D52 & 4  \\
\smallskip
29d & 23:57:48.035 &  -32:34:31.40 &0.33& 20.63   &0.10& 0.05  & 0.11& 20.19 &0.05& --0.69   &0.32 & 0.18  & --7.89    & --8.71 & (WN?)& D52 & 4 \\
30b & 23:57:48.376 &  -32:34:33.60 &0.32& 21.01   &0.09& --0.10& 0.14& 20.55 &0.05& --0.55   &0.09 & 0.18  & --7.51    & --8.06 & (WN?)& D55 & 4 \\
30a & 23:57:48.479 &  -32:34:33.48 &0.32& 20.40   &0.08& --0.18& 0.14& 19.98 &0.03& --0.23   &0.07 & 0.18  & --8.12    & --8.63 & (WN?)& D55 & 4\\
31b & 23:57:48.487 &  -32:37:01.57 &0.58& 20.31   &0.06& --0.41& 0.16& 19.51 &0.05& --0.42   &0.15 & 0.18  & --8.21    & --9.10 & (WN?)& D53 & 14 \\
31a & 23:57:48.535 &  -32:37:02.62 &0.59& 19.75   &0.05& --0.31& 0.15& 19.33 &0.04& --0.22   &0.05 & 0.18  & --8.77    & --9.28 & (WN?)& D53 & 14\\
\smallskip
31c & 23:57:48.665 &  -32:37:02.84 &0.59& 20.48   &0.09& --0.08& 0.14& 19.95 &0.07& --0.45   &0.07 & 0.18  & --8.04    & --8.66 & (WN?)& D53 & 14\\
32b & 23:57:48.560 &  -32:34:46.00 &0.24& --      &    & --    &     & 21.13 &0.05& --1.19   &0.08 & 0.18  &                        & --7.48 & (WC?) & D54 & 8 \\ 
32a & 23:57:48.644 &  -32:34:46.86 &0.23& 20.94   &0.10&   0.00&0.12 & 20.84 &0.03&  +0.13   &0.06 & 0.18  & --7.57    & --7.77 & (not WR?) & D54 & 8 \\
33  & 23:57:48.791 &  -32:34:35.37 &0.30& 19.56   &0.04& --0.16& 0.07& 19.15 &0.02& --0.25   &0.09 & 0.18   & --8.96   & --9.21 & (WN?) & D55 & 4 \\
34  & 23:57:48.816 &  -32:34:53.27 &0.19& 18.07   &0.01& --0.21& 0.01& 17.72 &0.01& --0.18   &0.24 & 0.32$^{3}$  & --10.88   & --11.39 & WN5--6:WC4 & GHR \#3 & 8\\
\smallskip
35  & 23:57:48.855 &  -32:34:32.86 &0.31& 21.57   &0.09& 0.01  & 0.46& 20.61 &0.04& --0.59   &0.16 & 0.09$^{3}$  & --6.66    & --7.66 & WC5--6 & D55 & 4 \\
36  & 23:57:48.891 &  -32:34:57.03 &0.17& 20.88   &0.02& 0.06  & 0.05& 20.20 &0.04& --0.90   &0.06 & 0.27  & $\phantom{.}$$<$--4.98 & --8.74 & WC6 & -- & 8 \\
37  & 23:57:48.958 &  -32:36:58.16 &0.56& 21.22   &0.08& 0.00  & 0.12& 20.43 &0.08& --0.74   &0.10 & 0.13  & --7.16    & --8.00 & WC4--6 & D53 & 14 \\
38  & 23:57:49.051 &  -32:33:32.36 &0.67& 22.90   &0.05&--0.07 & 0.06& 21.86 &0.03& --0.84   &0.06 & 0.18  & --5.62    & --6.75 & (WN?) & -- & 2\\
39  & 23:57:51.522 &  -32:35:01.33 &0.15& --      &    & --    &     & 20.77 &0.04& --0.44   &0.06 & 0.18$^{3}$  &                        & --7.84 & WN2--4 & D75 & 10 \\
\hline
\end{tabular}
\end{table*}%

\begin{table*}
{{\bf Table 2} (continued)} \\
\begin{tabular}{
l@{\hspace{2mm}}
c@{\hspace{2mm}}
c@{\hspace{2mm}}
c@{\hspace{2mm}}
r@{\hspace{1mm}}
l@{\hspace{2mm}}
r@{\hspace{1mm}}
l@{\hspace{2mm}}
r@{\hspace{1mm}}
l@{\hspace{2mm}}
r@{\hspace{1mm}}
l@{\hspace{2mm}}
c@{\hspace{2mm}}
c@{\hspace{2mm}}
c@{\hspace{2mm}}
c@{\hspace{2mm}}
c@{\hspace{2mm}}
c}
\hline\noalign{\smallskip}
  ID & RA       &   Dec    & r/$R_{25}$ & \multicolumn{2}{c}{m$_{V}$} & \multicolumn{2}{c}{m$_B$--m$_{V}$} & \multicolumn{2}{c}{m$_{4684}$}  & 
\multicolumn{2}{c}{m$_{4684}$--m$_{5100}$}  & E(B-V)  &  
M$_V$ & M$_{4684}$  &  Spectral      & H\,{\sc ii} & Finding \\ 
     & J2000    & J2000    & & mag     & $\pm$ & mag        & $\pm$ & mag     & $\pm$ & mag &$\pm$  & mag & mag & mag&  Type         &  region    & Chart \\
\noalign{\smallskip}\hline
40  & 23:57:52.025 &  -32:35:21.30 &0.12& 19.70  &0.04& --0.05 &0.39& 18.98 &0.05   & --0.56  &0.23 & 0.18  & --8.81    & --9.63 & (WN?) & D82 & 10 \\
41  & 23:57:53.045 &  -32:35:13.44 &0.17& 21.43  &0.05& --0.12 &0.09& 20.71 &0.03   & --0.48  &0.10 & 0.18  & --7.09    & --7.90 & (WN?) & D85 & 10 \\
42b & 23:57:53.540 &  -32:36:42.24 &0.54& 20.99  &0.09& 0.22   &0.10& 20.40 &0.04   & --0.57  &0.07 & 0.29  & --7.86    & --8.60 
& \multirow{2}{*}{$\sqsupset$WC4} & D88 & 16 \\
42a & 23:57:53.621 &  -32:36:42.98 &0.54& 20.65  &0.07& --0.03 &0.13& 20.34 &0.03   & --0.12  &0.17  & 0.29  & --8.20    & --8.66 &  & D88 & 16 \\
43  & 23:57:53.975 &  -32:35:15.25 &0.21& 23.29  &0.09&   0.16 &0.12& 21.61 &0.03   & --1.39  &0.12  & 0.18  &                        & --7.00 & WC4--6 & -- & 10 \\ 
\smallskip
44  & 23:57:54.090 &  -32:35:46.91 &0.28&  --     &   &  --    &    & 22.39 &0.11   & --1.64  &0.14  & 0.18  &                        & --6.22 & (WC?) & D91 & 10 \\
45  & 23:57:54.122 &  -32:35:27.83 &0.22& --      &   &  --    &    & 22.11 &0.06   &  --     &   & 0.18  &                        & --6.50 & WC4--6 & -- & 10 \\
46  & 23:57:54.325 &  -32:34:00.33 &0.51& 20.10  &0.07& --0.02 &0.08& 19.48 &0.02   & --0.78  &0.06  & 0.17  & --8.38    & --9.09 & WC4  & D95 & 7 \\
47  & 23:57:54.704 &  -32:35:30.35 &0.25& 21.04  &0.03& --0.32 &0.12& 19.92 &0.03   & --0.73  &0.06  & 0.17$^{3}$  & --7.44    & --8.65 & WC6 & D97 & 10\\
48  & 23:57:54.849 &  -32:35:29.64 &0.26& 20.02  &0.02& --0.05 &0.12& 19.56 &0.05   & --0.34  &0.08  & 0.18   & --8.50    & --9.05 & (WN?) & D97 & 10 \\
\smallskip
49  & 23:57:55.219 &  -32:33:55.78 &0.55& 22.12  &0.09& 0.30   &0.10& 21.31 &0.02   & --1.01  &0.09  & 0.18  & --6.40    & --7.30 & WN3 & D101 & 7 \\
50c & 23:57:55.838 &  -32:34:41.75 &0.36& --      &   & --     &    & 21.22 &0.04   & --      &      & 0.25$^{3}$   &                        & --6.41 & (WC?) & D106 & 13\\ 
50a & 23:57:55.880 &  -32:34:41.25 &0.36& 20.90  &0.04& --0.04 &0.11& 20.76 &0.03   & +0.10   &0.05  & 0.25$^{3}$   & --7.84    & --7.16 & WC4 & D106 & 13 \\
50b & 23:57:55.979 &  -32:34:43.09 &0.36& 21.22  &0.06& --0.17 &0.07& 20.88 &0.03   & --0.06  &0.05  & 0.25$^{3}$   & --7.52    & --7.98 & (not WR?) & D106 & 13 \\
51  & 23:57:56.310 &  -32:36:12.84 &0.47& 22.90  &0.07& 0.08   &0.12& 20.89 &0.05   & --1.19  &0.23  & 0.18  & --5.62    & --7.72 & (WC?) & D109 & 15 \\
\smallskip
52a & 23:57:56.875 &  -32:33:47.67 &0.62& 20.04  &0.06& --0.04 &0.08& 19.75 &0.03   & --0.20  &0.07  & 0.18  & --8.48    & --8.86 & (WN?) & D111 & 7 \\
52b & 23:57:56.919 &  -32:33:46.60 &0.63& 20.19  &0.05& 0.04   &0.08 & 20.05 &0.03&  --0.09   &0.22  & 0.18 & --8.33    & --8.56 & (not WR?)& D111 & 7 \\
53  & 23:57:57.171 &  -32:36:08.72 &0.49& 19.70  &0.05&--0.19  &0.08 & 19.60 &0.02&  --0.10   &0.35  & 0.18  & --8.82    & --9.01 & (not WR?) & D110 & 15 \\
54  & 23:57:57.205 &  -32:36:11.12 &0.50& 20.56  &0.06& 0.09   &0.12 & 20.21 &0.03&  --0.40   &0.03  & 0.18  & --7.96    & --8.40 & (WN?) & D110 & 15 \\
55  & 23:57:57.649 &  -32:35:39.16 &0.41& 23.84  &0.08& --0.40 &0.12 & 21.52 &0.02&   --      &      & 0.18  & --4.68    & --7.09 & WC5--6 & -- & 15 \\
\smallskip
56  & 23:57:57.867 &  -32:34:50.81 &0.41& 23.46  &0.06& --0.29 &0.10 & 21.79 &0.03&  --       &      & 0.18  & --5.06    & --6.82 & WN2--4 & -- & 13 \\
57  & 23:57:58.175 &  -32:35:40.98 &0.44& 21.96  &0.10& 0.12   &0.10 & 20.78 &0.03& --        &      & 0.18  & --6.56    & --7.83 & (WC?) & -- & 15 \\
58a & 23:57:58.548 &  -32:34:32.92 &0.48& 20.12  &0.07&--0.04  &0.14 & 19.71 &0.04& --0.12   & 0.51 & 0.18  & --8.40    & --8.90 & (not WR?) & D115 & 13  \\
58b & 23:57:58.602 &  -32:34:30.72 &0.49& 20.41  &0.07&--0.05  &0.11 & 20.14 &0.04& +0.01     & 0.09 & 0.18  & --8.11    & --8.47 & (not WR?) & D115 & 13  \\
59  & 23:57:58.635 &  -32:35:46.19 &0.47& 23.21  &0.06&--0.35  &0.08 & 21.50 &0.03& --        &      & 0.00  & --4.75    & --6.46 & WC6 & D116 & 15 \\
\smallskip
60  & 23:57:58.929 &  -32:36:46.90 &0.72& --      &   & --     &     & 19.02 & 0.03&--1.35    & 0.13 & 0.18  &                        & --9.59 & (WC?) & D117 & 18 \\
61a & 23:57:59.098 &  -32:36:50.32 &0.74& 19.72  &0.04& --0.06 &0.06 & 19.52 & 0.02&--0.09    &0.06  & 0.18  & --8.80    & --9.09 & (not WR?) & D118 & 18 \\
61b & 23:57:59.098 &  -32:36:48.96 &0.74& 19.96  &0.05& 0.00   &0.07 & 19.74 & 0.03&--0.01    &0.04  & 0.18  & --8.56    & --8.87 & (not WR?) & D118 & 18 \\
62  & 23:57:59.560 &  -32:36:42.42 &0.72& 22.16  &0.05&0.28    &0.06 & 21.41 & 0.02& --1.11   &0.07  & 0.07  & --6.02    & --6.81 & WN2--4 & --  & 18 \\
63  & 23:57:59.822 &  -32:34:23.47 &0.55&23.38$^{2}$&0.19$^{2}$&  &   & 22.04 & 0.04& --0.50   &0.10  & 0.18  & --5.32$^2$  & --6.57 & WN2--4 & --  & 13 \\
\smallskip
64  & 23:58:00.161 &  -32:35:33.11 &0.51& 20.03  &0.03& --0.26 &0.13 & 19.41 & 0.03& --0.09   &0.12 & 0.18  & --7.81    & --9.20 & (not WR?) & D121 & 15 \\
65  & 23:58:00.186 &  -32:33:22.03 &0.81& --      &   & --     &     & 22.47 & 0.08& --       &     & 0.18  &           & --6.14 & (WC?) & -- & 12 \\
66  & 23:58:00.222 &  -32:34:01.40 &0.64& --      &   & --     &     & 20.26 & 0.07& --0.27   &0.12 & 0.18  &                        & --8.35 & (WN?) & -- & 12 \\
67  & 23:58:00.239 &  -32:34:46.21 &0.52& 21.07  &0.06& 0.09   &0.11 & 20.66 & 0.03& --0.23   &0.08 & 0.18  & --7.45    & --7.95 & (WN?) & D127 & 13 \\
68  & 23:58:00.511 &  -32:34:10.54 &0.62& 20.52  &0.06& 0.03   &0.12 & 20.12 & 0.04& --0.21   &0.05 & 0.18  & --8.00    & -8.49 & (WN?) & D124 & 12 \\
\smallskip
69  & 23:58:00.718 &  -32:35:48.95 &0.57& 20.92  &0.05& --0.17 &0.05 & 20.06 & 0.04&--0.78    &0.06 & 0.15  & --7.49    & --8.43 & WN2--4 & -- & 15 \\
70b & 23:58:00.933 &  -32:33:56.59 &0.68& 20.63  &0.04&--0.21  &0.17 & 20.15 & 0.04&--0.14    &0.08 & 0.18  & --8.36 & --8.81  &
 \multirow{2}{*}{$\sqsupset$WN2--4} & D129 & 12\\
70a & 23:58:00.984 &  -32:33:56.22 &0.69& 20.16  &0.04&--0.12  &0.14 & 19.80 & 0.02& --0.18   &0.16 & 0.18  & --7.89    & --8.46 &    & D129 & 12 \\
71  & 23:58:01.194 &  -32:33:37.19 &0.77& 20.31  &0.06& --0.06 &0.15 & 20.04 &0.05 & --0.37   &0.08 & 0.18  & --8.21    & --8.57 & (WN?) & D130 & 12 \\
72  & 23:58:01.720 &  -32:33:46.37 &0.75& 22.37  &0.10& --0.23 &0.10 & 21.17 & 0.03& --0.77   &0.45 & 0.13  &                        & --7.24  & WN2--4 & D131 & 12 \\
\smallskip
73a & 23:58:06.631 &  -32:34:53.44 &0.79& n/a     &   & n/a    &      & 20.21 &0.08& --0.25   &0.40 & 0.12  & n/a       & --8.18 & WC4 & D132 & 17 \\
73b & 23:58:06.666 &  -32:34:51.68 &0.80& n/a     &   & n/a    &      & 19.53 &0.04& --1.37   &0.05 & 0.12  & n/a       & --8.86 & (WC?) & D132 & 17 \\
74a & 23:58:06.898 &  -32:34:56.17 &0.81& n/a     &   & n/a    &      & 20.82 &0.04& --0.39   &0.06 & 0.18  & n/a       & --7.79 & (WN?) & D132 & 17 \\
74b & 23:58:07.005 &  -32:34:55.94 &0.81& n/a     &   & n/a    &      & 21.18 &0.03& --0.12   &0.03 & 0.18  & n/a       & --7.43 & (not WR?) & D132 & 17 \\
\hline
Q2358--32  & 23:58:02.881 &  -32:36:14.03 & &20.79&0.01 &  0.20  &0.02  & 20.38 &0.01& --0.62   &0.08 &        &          &  & QSO  & -- & 19 \\
\hline
\multicolumn{4}{l}{\footnotesize{$^1$m$_{5100}$ photometry}} \\
\multicolumn{4}{l}{\footnotesize{$^2$m$_{B}$ photometry}}\\
\multicolumn{4}{l}{\footnotesize{$^3$Upper limit  since weak nebular H$\beta$ emission}}\\
\end{tabular}
\end{table*}%

\subsection{Spectroscopy, Flux Calibration \& Slit losses}

VLT/FORS1 Multi Object Spectroscopy (MOS) was undertaken in July 2008
with the standard resolution collimator and the 300V grism (dispersion
of $\sim$3\,$\AA$ pixel$^{-1}$) at a central wavelength of
$\lambda$~5850. Spectra were obtained using 0.8~$\arcsec$
slits resulting in a spectral resolution of $\sim$8\,$\AA$, with seeing
conditions between 1 and 1.25~$\arcsec$. The typical wavelength range
of each slit was 3600-9200\,$\AA$ however fringing effects made data
longward of 7000\,$\AA$ unreliable.

Each of the MOS masks that were designed contained 19 slits with
lengths ranging from 22 to 26\,$\arcsec$ and used exposure times of 1650
seconds. Due to the spatial distribution of our candidates only
$\sim$13 slits per mask could be placed on our primary sources. The
remaining slits were used to obtain either multiple spectra of a
candidate or a HII region. Out of the 74 candidates we obtained
spectra of 39 of our WR candidates in Table~\ref{sources}, some
containing multiple sub-regions. 

Spectroscopic data were bias subtracted and extracted using
\textsc{iraf} whilst wavelength and flux calibration were achieved
with \textsc{figaro}. Wavelength calibration was achieved from an He
HgCdAr arc lamp. Spectrophotometric standard stars LTT 377 (masks \#1
\& \#2) and LTT 1788 (mask \#3) \citep{Hamuy1994, Hamuy1992} were used
to produce a relative flux calibration whilst an absolute flux
calibration was determined from comparison between the synthetic
magnitudes and photometry. The sythetic magnitudes were obtained from
the convolution of spectra with appropriate filters. This factor
corrects our spectroscopic magnitudes for slit losses, which we
determine to be 1.49$\pm$0.3.

\section{Nebular Analysis}

In this section we determine the nebular properties of individual H\,{\sc ii}
regions in NGC~7793 and determine an average extinction and
metallicity. We compare our results to previous work, and discuss the
presence of a metallicity gradient in NGC~7793.

\begin{table*}

\caption{A sample of observed, F$_{\lambda}$, and dereddened, I$_{\lambda}$
  nebular fluxes of H\,{\sc ii} regions in NGC~7793, relative to
  H$\beta$. The final row lists H$\beta$ fluxes in units of $\times
  10^{-15}$ erg s$^{-1}$cm$^{-2}$. The complete version of this table is available online.}
\begin{tabular}{cccccccccccccc}
\hline
$\lambda$($\AA$) & ID & \multicolumn{2}{c}{1} &\multicolumn{2}{c}{3} & \multicolumn{2}{c}{4} &\multicolumn{2}{c}{5} & \multicolumn{2}{c}{7a--b} & \multicolumn{2}{c}{9}  \\

    &    & $F_{\lambda}$ & $I_{\lambda}$ & $F_{\lambda}$ & $I_{\lambda}$   & $F_{\lambda}$ & $I_{\lambda}$ & $F_{\lambda}$ & $I_{\lambda}$ & $F_{\lambda}$ & $I_{\lambda}$ & $F_{\lambda}$ & $I_{\lambda}$ \\
\hline 
3727  & [O\,{\sc ii}]     & 254 & 306 & 251 & 287 & 547 & 612 & 171 & 199 & 296 & 335 & 154 & 154  \\ 
4343  & H$\gamma$         & 38  & 42  & 44  & 47  & 43  &  53 & 40  & 43  & 41  & 43  & 53  & 53    \\  
4861  & H$\beta$          & 100 & 100 & 100 & 100 & 100 & 100 & 100 & 100 & 100 & 100 & 100 & 100   \\ 
4959  & [O\,{\sc iii}]    & 45  & 44  & 74  & 73  & 73  & 72  & 92  & 91  & 29  & 29  & 22  & 22    \\  
5007  & [O\,{\sc iii}]    & 138 & 135 & 221 & 217 & 239 & 236 & 272 & 266 & 86  & 84  & 66  & 66    \\  
6563  & H$\alpha$         & 365 & 290 & 340 & 288 & 331 & 288 & 352 & 290 & 338 & 289 & 232 & 192  \\ 
6583  & [N\,{\sc ii}]     & 40  & 31  & 34  & 29  & 72  & 63  & 21  & 17  & 56  & 48  & 41  & 41   \\ 
6716  & [S\,{\sc ii}]     & 27  & 21  & 28  & 23  & 42  & 36  & 15  & 12  & 38  & 32  & 18  & 18   \\ 
6731  & [S\,{\sc ii}]     & 19  & 15  & 20  & 17  & 34  & 29  & 10  & 8   & 27  & 23  & 7   & 7    \\ 
\hline
4681  & H$\beta$  & 0.901 & 1.85 & 2.92 & 4.89 & 0.104 & 0.161 & 2.08 & 3.80 & 4.39 & 7.15 & 0.308 & 0.308  \\ 
\hline

\end{tabular}
\label{fluxes_sample}
\end{table*}

\subsection{Interstellar Extinction}

\begin{figure}
   \centering
   \includegraphics[width=0.7\columnwidth,angle=-90]{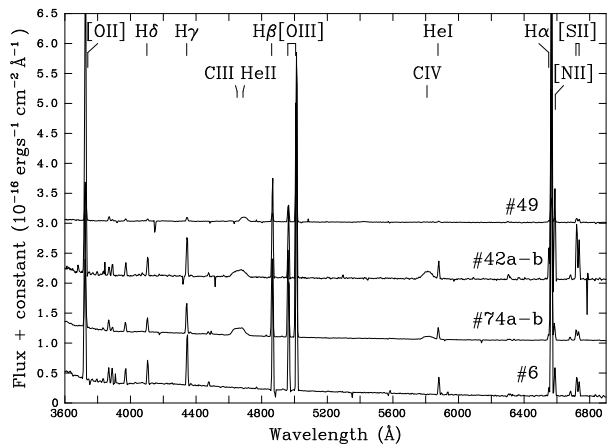}
     \caption{VLT/FORS spectroscopy showing examples of nebular and WR
       line features for sources in NGC~7793. In sources \#42 and 72
       the C\,{\sc iii}, He\,{\sc ii} and C\,{\sc iv} emission are WR
       features. For clarity spectra are successively offset by
       1x10$^{-16}$erg s$^{-1}$cm$^{-2}\AA^{-1}$.}
                   \label{nebular}
   \end{figure}

70\% of the 42 MOS spectra reveal nebular Balmer line emission which
we can use to derive the interstellar extinction of the
region. Examples are presented in Figure~\ref{nebular}. We used the
Emission Line Fitting (\textsc{elf}) routine within the
\textsc{starlink} package \textsc{dipso} to fit Gaussian profiles to
the Balmer lines in the extracted spectra. Using Case B recombination
theory \citep{Hummer1987} we assumed an electron density of
n$_{e}$$\sim$100 cm$^{-3}$ and temperature T$_{e}$$\sim$10$^{4}$K
together with a standard Galactic extinction law \citep{Seaton1979} to
estimate the interstellar extinction from the observed
H$\alpha$/H$\beta$ line ratio. Observed and extinction corrected
nebular line fluxes are shown for six H\,{\sc ii} regions in Table
\ref{fluxes_sample}, while a complete catalogue of the 29 H\,{\sc ii}
regions (including the nucleus) is avaliable online.

We note that we do not formally correct
for the underlying stellar Balmer line absorption, quantified by
\citet{McCall1982} to be of order 2$\AA$ for H$\alpha$ and
H$\beta$. This results in a larger percentage error for nebular lines
with a small equivalent width (EW). \citet{Mazzarella&Boroson1993}
conclude that the underlying absorption causes their extinction to be
overestimated by 15-20\%.  Our nebular lines have an EW(H$\beta$) in
the range 10-100$\AA$ so we estimate errors of $\sim$2-20\% (0.01-0.1
mag) on our E(B-V) values. Sources with larger errors are highlighted
in Table~\ref{sources} with quoted E(B-V) values representing upper
limits.


Excluding a few outliers with unphysical extinction measurements,
derived values range from the Galactic foreground
extinction of E(B-V)\,=\,0.019 mag \citep{Schlegel1998} to 0.319
magnitudes. The average is E(B-V)\,=\,0.179$\pm$0.024 magnitudes. Our
values are much lower than those found by \citet{Chun1983} who find
extinctions in the range E(B-V)\,=\,0.56-0.88 mag. However, they are more
consistent with \citet{Webster1983} who measure c(H$\beta$)=
E(B-V)/0.7\,=\,0.23--0.61. Unfortunately there are no H\,{\sc ii} regions
common to all three samples so a direct comparison cannot be made.

\subsection{Metallicity}
Elemental abundances are best determined from the weak line method
described in \citet{Osterbrock1989}. This method relies on accurate
determination of the temperature of the region from the weak [O\,{\sc
    iii}]\,$\lambda$4363 line and is preferable to strong line
methods. Unfortunately, the [O\,{\sc iii}]\,$\lambda$4363 line is
marginally detected in only four of the H\,{\sc ii} regions observed
in our MOS spectra, and moreover the detection is below the 3$\sigma$
level. As a result the errors associated with the derived temperatures
are too high to allow reliable abundance determinations and so we
resort to strong line methods.

\begin{figure}
   \centering
   \includegraphics[width=0.7\columnwidth,angle=-90]{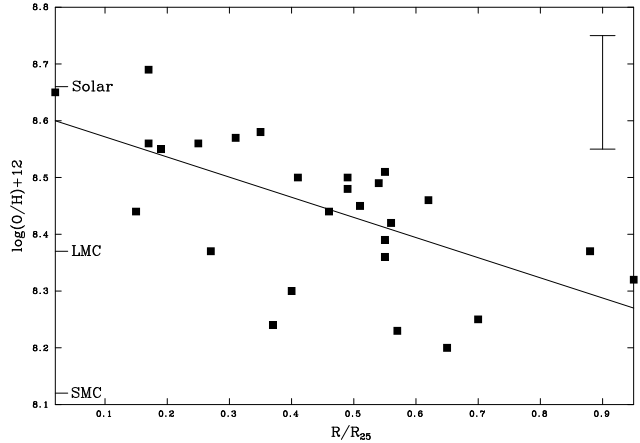}
     \caption{Comparison of the metallicity of H\,{\sc ii} regions within
       NGC~7793 relative to their position from the centre of the
       galaxy. Typical error bars are shown (0.2 dex).}
                   \label{metalgradient}
   \end{figure}

By calculating both the I([N\,{\sc ii}]\,$\lambda$6584)/I(H$\alpha$)
and I([O\,{\sc iii}]\,$\lambda$5007)/I(H$\beta$) ratios we can use the
N2 and O3N2 indicies derived in \citet{Pettini&Pagel2004} to determine
the abundance of H\,{\sc ii} regions within NGC~7793 to $\pm$0.2
dex. Metallicities of the H\,{\sc ii} regions ranges from log(O/H)+12
= 8.19--8.69 with an average value of log(O/H)+12\,=\,8.44$\pm$0.24
(See Table \ref{abundances}). This is consistent with previous
estimates of log(O/H)+12\,=\,8.54 for NGC~7793 by
\citet{Pilyugin2004}, but is somewhat lower than log(O/H)+12\,=\,8.7
quoted in \citet{O'Halloran2006}. This difference may be explained by
the presence of metallicity gradient of approximately
\begin{center}
\begin{displaymath}\log\frac{O}{H} + 12 = (8.61\pm0.05) - (0.36\pm0.01)\frac{r}{R_{25}} \end{displaymath}
\end{center}
as shown in Figure \ref{metalgradient}. \citet{Webster1983} also found
evidence of a metallicity gradient in NGC~7793 with
12+log(O/H)\,=\,8.23--8.96.

\begin{table}
\caption{Deprojected distances from the centre of the galaxy, N2 and O3N2 derived
  metallicities for H\,{\sc ii} regions within NGC~7793. The error on the average
  value is $\pm$0.24 dex.}
\begin{tabular}{c@{\hspace{1mm}}c@{\hspace{1mm}}c@{\hspace{1mm}}c@{\hspace{1mm}}c@{\hspace{1mm}}c@{\hspace{1mm}}c}
\hline
Source & r/R$_{25}$ & \underline{I([NII])} & log(O/H) & \underline{I([OIII])} & log(O/H) & log(O/H) \\
   ID  &           &   I(H$\alpha$)         & + 12 $^{1}$ &         I(H$\beta$)   & + 12 $^{2}$   & + 12 $_{mean}$ \\   
\hline
1      & 0.88 & 0.109 & 8.35 & 1.35 & 8.38 & 8.37 \\
3      & 0.95 & 0.100 & 8.33 & 2.18 & 8.30 & 8.32 \\
4      & 0.62 & 0.218 & 8.52 & 2.36 & 8.40 & 8.46 \\
6      & 0.65 & 0.059 & 8.20 & 2.66 & 8.20 & 8.20  \\
7ab    & 0.49 & 0.165 & 8.45 & 0.84 & 8.50 & 8.48  \\
9      & 0.55 & 0.176 & 8.47 & 0.64 & 8.55 & 8.51 \\
10     & 0.40 & 0.115 & 8.36 & 3.82 & 8.25 & 8.30 \\
11a-c  & 0.46 & 0.164 & 8.45 & 1.46 & 8.43 & 8.44  \\
14     & 0.37 & 0.067 & 8.23 & 2.24 & 8.24 & 8.24 \\
16     & 0.70 & 0.088 & 8.30 & 4.18 & 8.19 & 8.25  \\
20     & 0.27 & 0.399 & 8.67 & --   & --   & 8.67 \\
22a-c  & 0.27 & 0.116 & 8.37 & 1.56 & 8.37 & 8.37  \\
23     & 0.55 & 0.166 & 8.46 & 3.26 & 8.32 & 8.39  \\
26     & 0.17 & 0.459 & 8.71 & 0.68 & 8.68 & 8.69  \\
27     & 0.49 & 0.220 & 8.53 & 1.42 & 8.47 & 8.50  \\
34     & 0.19 & 0.256 & 8.56 & 0.97 & 8.55 & 8.55  \\
35     & 0.31 & 0.280 & 8.58 & 0.94 & 8.56 & 8.57  \\
36     & 0.17 & 0.233 & 8.54 & 0.74 & 8.57 & 8.56  \\
37     & 0.56 & 0.145 & 8.42 & 1.37 & 8.42 & 8.42  \\
39     & 0.15 & 0.158 & 8.44 & --   & --   & 8.44  \\
42ab   & 0.54 & 0.234 & 8.54 & 1.80 & 8.45 & 8.49  \\
46     & 0.51 & 0.175 & 8.47 & 1.42 & 8.44 & 8.45  \\
47     & 0.25 & 0.231 & 8.54 & 0.70 & 8.58 & 8.56  \\ 
49     & 0.55 & 0.178 & 8.47 & 5.45 & 8.25 & 8.36  \\
50a-c  & 0.35 & 0.548 & 8.56 & 1.28 & 8.61 & 8.58  \\
62     & 0.72 & 0.139 & 8.41 & 3.33 & 8.29 & 8.35  \\
69     & 0.57 & 0.082 & 8.28 & 4.11 & 8.19 & 8.23  \\
\hline
Average &     & 0.19  &       &      &      & 8.44$\pm$0.24 \\
\hline
\multicolumn{7}{l}{$^{1}$ \footnotesize{N2}} \\
\multicolumn{7}{l}{$^{2}$ \footnotesize{O3N2}} \\

\end{tabular}
\label{abundances}
\end{table}

\section{The Wolf-Rayet Population of NGC~7793}
Wolf-Rayet stars are relatively straightforward to identify and
classify from optical spectra due to their strong, broad emission
lines. Nitrogen-rich (WN) stars are dominated by He\,{\sc
  ii}\,$\lambda$4686, while carbon-rich (WC) stars are dominated by
C\,{\sc iii}\,$\lambda$4650 and C\,{\sc iv}\,$\lambda$5801--12 and
oxygen rich (WO) stars by O\,{\sc vi}\,$\lambda$3811--34. Individual
WC stars possess much larger equivalent widths than WN stars,
therefore, allowing for dilution by neighbouring stars (see Section
5.2), this can result in an observational bias towards WC stars
\citep{Massey&Johnson1998, Crowther2003}.

From our 74 candidate WR regions regions identified from the continuum
subtracted $\lambda$\,4684 image we obtained spectra of 39. For these
sources the He\,{\sc ii}/C\,{\sc iii} excess ranged from +0.1 to --1.5
mag. From the 39 spectra, 33 revealed WR features ($\geq$ 3$\sigma$),
while 6 showed solely nebular lines. This 85\% detection rate is
consistent with spectroscopy of NGC~1313 by \citet{Hadfield2007} who
found 70 of their 82 candidates to exhibit WR features. The sources in
NGC~7793 that do not contain WR emission have m$_{4684}$-m$_{5100}$=
--0.22$\pm$0.1 mag. It is likely that these false candidates arise
from the formal photometric errors of 0.1-0.2 mag (see Section
2.3). We note that from the subregions found using higher spatial
resolution broad-band imaging we obtained spectra in 43 cases, 35 of
which have WR features.

\begin{figure}
   \centering
   \includegraphics[width=0.7\columnwidth, angle=-90]{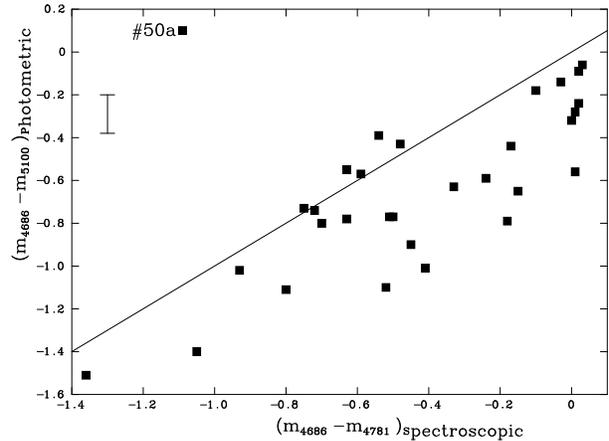}
     \caption{Plot of spectroscopic magnitudes against photometric
       magnitudes for sources in NGC 7793 of which we have
       spectra. Outliers indicate unreliable magnitudes determined
       from photometry. Photometric errors are shown, spectroscopic
       errors are not due to large variance.}
                   \label{outliers}
   \end{figure}

Figure \ref{outliers} compares the photometric He\,{\sc ii} excess to
the spectroscopic excess for all of our spectroscopic sources. The
spectroscopic excess was calculated from the convolution of sythetic
spectra with appropriate narrow band filters. Outliers from the trend
indicate that the magnitude derived from photometry is
unreliable. There is reasonable agreement with the exception of
sources \# 50a owing to severe crowding.

Using the \textsc{dipso} emission line fitting (\textsc{elf}) routine,
the flux and FWHM of the emission lines could be measured these are
presented in Table \ref{WR}. Typical errors on the line flux
measurements were $\sim$5\% for strong lines such as
HeII\,$\lambda$4686 and CIV\,$\lambda$5808, but were significantly
higher, $\sim$20\%, for weaker lines such as NV\,$\lambda$4603 and
HeII $\lambda$5411. Given the similar metallicity of NGC 7793 to the
LMC we use line fluxes of LMC WR stars found in \citet{Crowther2006}
as a comparison to assess the number of WR stars in each source. A
larger WR population would be inferred from SMC WR line luminosities,
while the reverse would be the case using Milky Way templates.

\begin{figure*}
  \begin{center}
   \includegraphics[width=1.5\columnwidth,angle=-90]{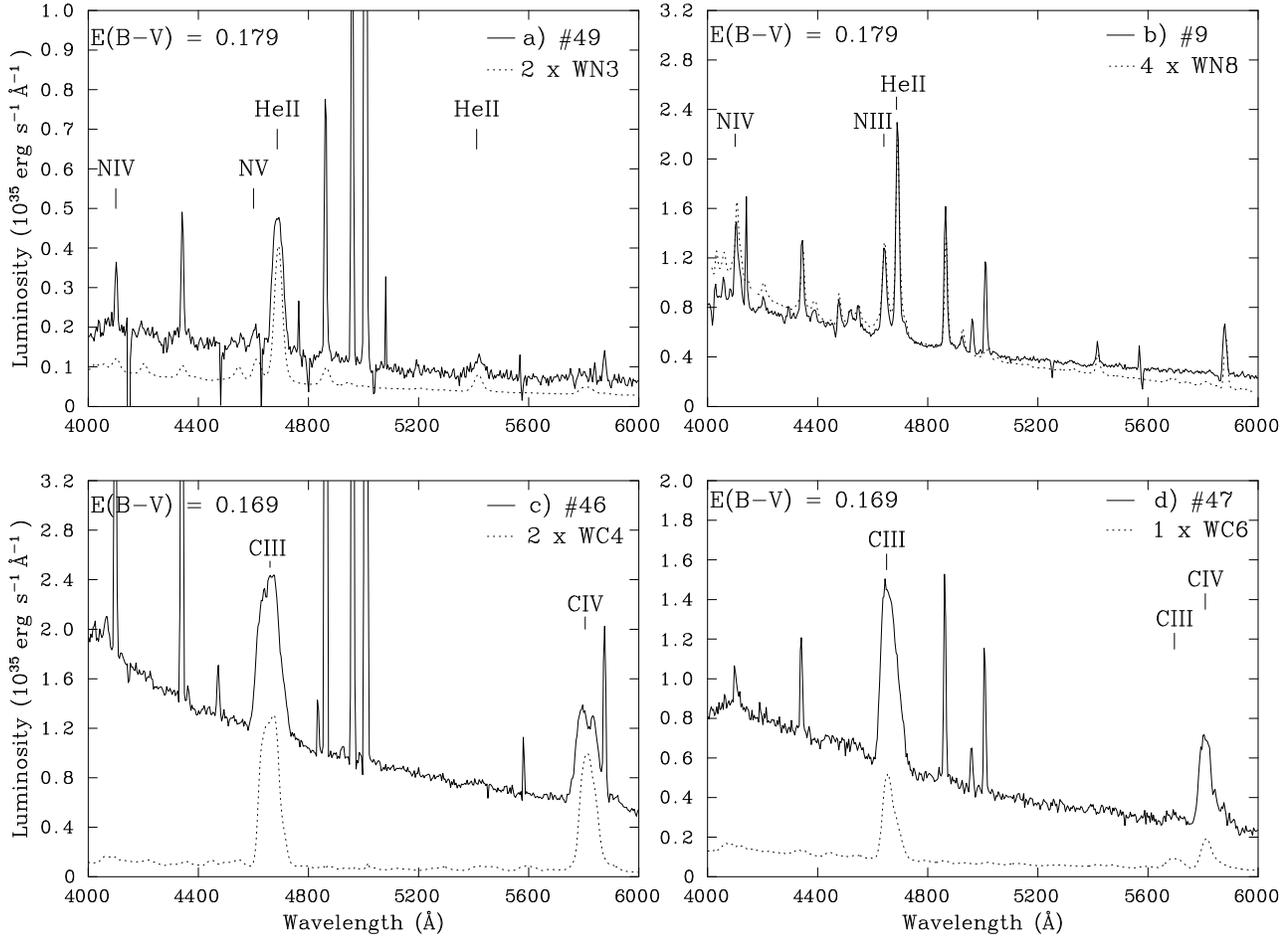}
     \caption{Extinction and distance corrected spectra of WN stars [(a) and (b)] and WC
       stars [(c) and (d)] in NGC~7793 (solid lines) compared to
       templates of LMC WR stars (dashed lines)}
                   \label{wr_spectra}
\end{center}
   \end{figure*}

\subsection{WN Subtypes}
Using the classification scheme derived by \citet{Smith1996} we can
divide WN stars into early (WNE, WN2--4), mid (WN5--6) and late (WNL,
WN7--9) categories, which are dominated by N\,{\sc
  v}\,$\lambda$4603--20, N\,{\sc iv}\,$\lambda$4058 + N\,{\sc
  iii}\,$\lambda$4636--4641 and N\,{\sc iii}\,$\lambda$4634--41
respectively. If only HeII\,$\lambda$4686 was detected we assumed a
WNE subtype based on the broad line emission (FWHM(WNE)=20--60$\AA$),
relative to narrower emission line (FWHM(WNL)$<$20$\AA$) seen in
late-type WR stars \citep{Crowther&Smith1997}. From comparison with WR
luminosities in the LMC \citep{Crowther2006} we can estimate the
number of WN stars that are responsible for the HeII\,$\lambda$4686
emission. Figure \ref{wr_spectra} a) and b) show examples of WNE and
WNL stars in NGC~7793, respectively, together with the spectra of
template stars in the LMC. From all our spectra we estimate $\sim$27
WN stars present within 13 regions (14 subregions), of which the
majority ($>$80\%) are WNE stars, which is a similar fraction to the
LMC \citep{Breysacher1999}.

\subsection{WC Subtypes}
The larger equivalent width of emission lines in WC stars make them
easier to detect compared to WN stars
\citep{Massey&Johnson1998}. \citet{Smith1990} produced a quantitative
classification scheme for WC stars based on the de-reddened line flux
ratio of C\,{\sc iv}\,$\lambda$5808/C\,{\sc iii}\,$\lambda$5696 and
C\,{\sc iii}\,$\lambda$5696/O\,{\sc v}\,$\lambda$5590. This was
further refined by \citet{Crowther1998} whose subtype classification
we use here. For sources \#36, \#47 and \#59 we determine a WC6
subtype, for other sources where C\,{\sc iii}\,$\lambda$5696 is
present the detection is below the 3~$\sigma$ level.  Hence we cannot
determine a more accurate subtype than WC4--6 in such cases. No late
type WC stars are found in NGC~7793, again consistent with the LMC
\citep{Breysacher1999} and other metal-poor galaxies.

Figures \ref{wr_spectra} c) and d) show examples of a WC4 and WC6
spectra respectively, and are again compared to LMC template WR stars
from \citet{Crowther2006}. The WC stars lie within bright regions
arising from unrelated OB stars in the source. We estimate a total of
$\sim$25 WC stars from 20 sources spectroscopically observed.

\subsection{Composite WN and WC spectra}

\begin{figure}
   \centering
   \includegraphics[width=0.7\columnwidth,angle=-90]{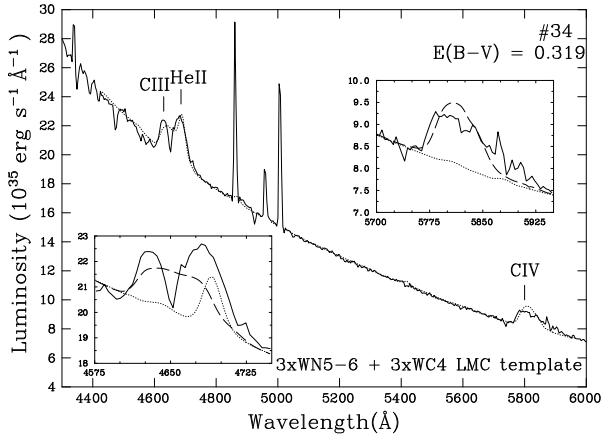}
     \caption{Source \#34 in NGC 7793. Solid line is VLT/FORS
       data. Dotted line (main plot) is LMC template of 3 WN5-6 stars
       and 3 WC4 stars. Inset show contributions from WC4 stars
       (dashed line) and WN5-6 stars (dotted line).}
                   \label{a29}
   \end{figure}

Of the 33 regions displaying WR signatures only source \#34 displays
the spectral signature of both WN and WC stars. In view of the large
H$\alpha$ flux of this region, which corresponds to $\sim$20 O7V stars
(GHR \#3; Table \ref{HIIregions}), we shall assume the WN stars
present are WN5-6 stars since this subtype is seen to dominate in
young bright H\,{\sc ii} regions \citep{Crowther&Dessart1998}. From
spectral line fitting, using LMC template stars we estimate that 3
WN5--6 and 3 WC4 stars are located in this region as shown in Figure
\ref{a29}. In total, from spectroscopy we find 52 WR stars in
NGC~7793, 10\% of which are located within the region of candidate
\#34 (See Section 8).

\subsection{WC Line Widths}

\begin{figure}
   \begin{center}
   \includegraphics[width=0.7\columnwidth,angle=-90]{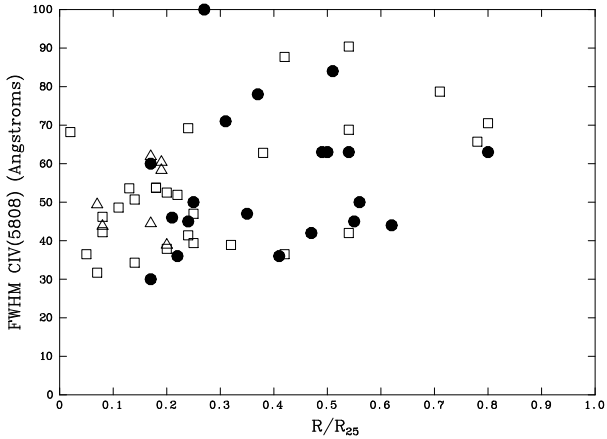}
     \caption{FWHM C\,{\sc iv}\,$\lambda$5808 increases with
         r/R$_{25}$ for WC4-6 stars in NGC 7793 and M33. NGC 7793 data
         (circles) are plotted with unpublished WHT/WYFFOS (empty
         squares) and CFHT/MOS data (empty triangles) for M33
         \citep{Abbott2004}.}
     \label{fwhm_wc}
     \end{center}
 \end{figure}

\citet{Schild1990} have studied the WC content of M33 revealing a
correlation between line width (C\,{\sc iv}\,$\lambda$5808 FWHM) and
galactocentric distance (r/R$_{25}$), supported by results of CFHT
spectroscopy in \citet{Abbott2004}. In Figure \ref{fwhm_wc} we
re-produce figure 6 from \citet{Abbott2004} together with unpublished
M33 WHT/WYFFOS data taken in August 1998, plus our own NGC~7793
observations. Indeed, there is a deficit of broad-lined WC4--6 stars
at low galactocentric distances.

The combined sample of WC stars, in M33 and NGC~7793, possess stronger
winds in the moderately metal-rich inner, compared to the metal-poor
outer regions. \citet{Crowther2002} argued that the difference between
WC subtypes in metal-poor and metal-rich galaxies was
metallicity-dependent winds. Metal-poor regions in galaxies exhibit
(broad-lined) WC4 stars while metal-rich regions possess
(narrow-lined) WC8--9 stars. In between these extremes the dominant
subtype would naturally be $\sim$WC6 subtypes as is the case for the
inner regions of NGC~7793 and M33. We would expect broad line sources
to have fast winds which is not consistent with their metal-poor
location. The physical explanantion behind this apparent
anti-correlation of metallicity and wind velocity is not understood.

\begin{figure}
   \centering
   \includegraphics[width=0.7\columnwidth,angle=-90]{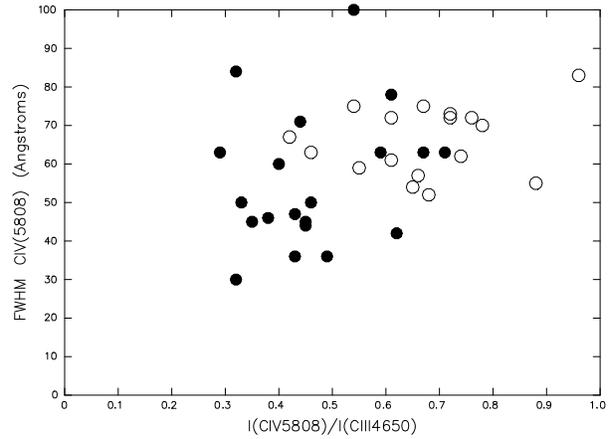}
     \caption{FWHM C\,{\sc iv}\,$\lambda$5808 increases with
       I(C\,{\sc iv}$\lambda$5808/C\,{\sc iii}\,$\lambda$4650) line
       ratio for WC4--6 stars. NGC~7793 data (filled circles) is
       compared to LMC WC4 data (empty circles)}
                   \label{fwhm_ratio}
   \end{figure}

For the WCE stars, there appears to be a correlation between the FWHM
of C\,{\sc iv}\,$\lambda$5808 and the relative strength of the
C\,{\sc iv}\,$\lambda$5808/C\,{\sc iii}\,$\lambda$4650 lines, with
narrower lines having weaker C\,{\sc iv}\,$\lambda$5808. The WC4--6
stars in NGC~7793 appear to span a larger range than the WC4 stars in
the LMC \citep{Crowther2006}. This is presented in Figure
\ref{fwhm_ratio} and would be a natural consequence of somewhat later
(WC5--6) subtypes for most NGC~7793 WC stars.

\begin{table*}
\caption{WR features, observed flux (F$_{\lambda}$) and extinction
  corrected luminosities (L$_{\lambda}$) based on a distance of
  3.91Mpc and E(B-V) values given in Table 2. Values in parentheses
  indicates a less secure detection ($<$3$\sigma$). Number of WR
  stars are based on the line luminosities for one WR star from
  \citet{Crowther2006}.}
\begin{tabular*}{0.75\textwidth}{ccccccccccc}
\hline
        & \multicolumn{6}{c}{F$_{\lambda}$($\times 10^{-16}$ erg s$^{-1}$ cm$^{-2}$)}  & \multicolumn{2}{c}{L$_{\lambda}$($\times 10^{36}$ erg s$^{-1}$ )} &   \\ 
Source  & F(N\,{\sc v}/N\,{\sc iii}) & F(C\,{\sc iii})   & F(He\,{\sc ii})  & F(He\,{\sc ii})  & F(C\,{\sc iii})  & F(C\,{\sc iv})   & L(He\,{\sc ii}) &  L(C\,{\sc iv}) & WR Subtype &  N(WR)   \\
        & 4603-4641               & 4647-4651      & 4686          &  5411         & 5696  & 5808          &    4686         &   5808    &        \\  
\hline
\# 4  & --                      & 8.95           &  --           & --           &  --    & 4.54          &     --          &  1.17    &  WC4--6 & 1  \\
\# 8  &                         & --             & 2.46          & --           & --     &  --           &  0.84           &  --      &  WN10 & 1 \\
\# 9  &    4.07                 & --             & 8.00          & 0.70         & --     &  --           &  2.72           &   --     &  WN8 & 4 \\
\# 10 &  1.52                   & --             & 12.7          & 1.78         & --     & --            &  4.14           &   --     &  WN2--4b  & 4 \\
\# 14  & --                      & 5.73           & --            & --           & --     & 4.51          & --              & 1.82     &  WC4 & 1 \\
\# 16  & --                      & --             & 6.03          & --           & --     & --            & 2.77            & --       & WN2--4  & 3 \\
\# 20  & --                      & 12.04          & --            & --           & --     & 7.56          & --              & 2.22     &  WC4 & 1 \\
\# 23  & --                      & 11.59          & --            & --           & --     & 5.81          & --              & 1.49     &  WC4 & 1 \\
\# 24  & --                      & 11.2           & --            & --           & (0.84) & 3.46          & --              & 0.97     &  WC4  & 1 \\
\# 25  & --                      & 2.27           & --            & --           & --     & 1.55          & --              & 0.46     & WC4  & 1\\
\# 26  & --                      & 4.14           & --            & --           & (0.30) & 1.52          & --              & 0.43     & WC4--6 & 1 \\
\# 27  & --                      & 4.53           & --            & --           & --     & 3.77          & --              & 1.11     & WC4 & 1 \\
\# 34 & (5.87)                  &                & 19.0          & --           & --     & 19.8          & 10.5            & 8.43     &  WN5--6:WC4 & 3:3 \\
\# 35  & --                      & 4.43           & --            & --           & --     & 2.08          & --              & 0.48     &  WC5--6 & 1  \\
\# 36  & --                      & 14.21          & --            & --           & 0.98   & 7.15          & --              & 2.68     & WC6 & 1 \\
\# 37  & --                      & 15.43          & --            & --           & (0.30) & 7.87          & --              & 2.03     &  WC4--6 & 1 \\
\# 39  & --                      &  --            & 3.17          & --           &  --    & --            & 1.08            &  --      & WN2--4 & 2 \\
\# 42ab & --                      & 13.90          & --            & --           & --     & 11.7          & --              & 4.59     &  WC4  & 2 \\
\# 43  & --                      & 7.25           & --            & --           & (0.20) & 3.25          & --              & 0.95     &  WC4--6 & 1 \\
\# 45  & --                      & 4.38           & --            & --           & (0.37) & 2.17          &--               & 0.64     & WC4--6 & 1 \\
\# 46 & --                   & 35.56          & --            & --           & --     & 12.7          & --              & 3.22     &  WC4 & 2 \\
\# 47  & --                      & 18.94          & --            & --           & 0.38   & 7.23          & --              & 2.03     &  WC6 & 1 \\
\# 49   & 0.62                    & --             & 4.73          & 6.06         & --     & --            & 1.61            &  --      &  WN3 & 2 \\
\# 50a  & --                      & 6.85           & --            & --           & --     & 3.62          & --              & 1.28     &  WC4 & 1 \\
\# 55  & --                      & 6.57           & 2.35          & --           & --     & 3.67          & --              & 1.08     & WC5--6 & 1 \\
\# 56   & 1.39                    & --             & 1.27          & 0.66         & --     & --            & 0.43            & --       & WN2--4 & 1 \\
\# 59  & --                      & 5.76           & --            & --           & 0.33   & 4.14          & --              & 1.21     &  WC6 & 1 \\
\# 62  & --                      & --             & 6.10          & 0.69         & --     & --            & 1.43            & --       &  WN2--4 & 2 \\
\# 63   & --                      & --             & 2.65          & 0.34         &   --   & 0.87          & 0.18            & 0.26     &  WN2--4 & 1 \\
\# 69  & 0.87                    & --             & 4.77          & --           & --     & --            & 1.44            &   --     &  WN2--4 & 2\\
\# 70ab   & --                      & --             & 1.56          & --           & --     & --            & 0.53            & --       & WN2--4 & 1 \\
\# 72   & 0.36                    & --             & 1.33          & --           & --     & --            & 0.74            & --       & WN2--4 & 1 \\
\# 73a  & --                      & 13.93          & --            & --           & --     & 4.44          & --              & 1.11     &  WC4 & 1  \\
\hline
\end{tabular*}
\label{WR}
\end{table*}

\section{Comparison of VLT and HST data}

In this section we consider the advantages of space-based observations
in identifying WR sources in galaxies beyond the Local Group. We
consider the case of the LMC to assess what fraction of known LMC
WR stars would have been detected at the distance of NGC~7793.

\subsection{Spatial Resolution}
\begin{figure}
   \centering 

\subfigure[]{
     \includegraphics[width=0.9\columnwidth]{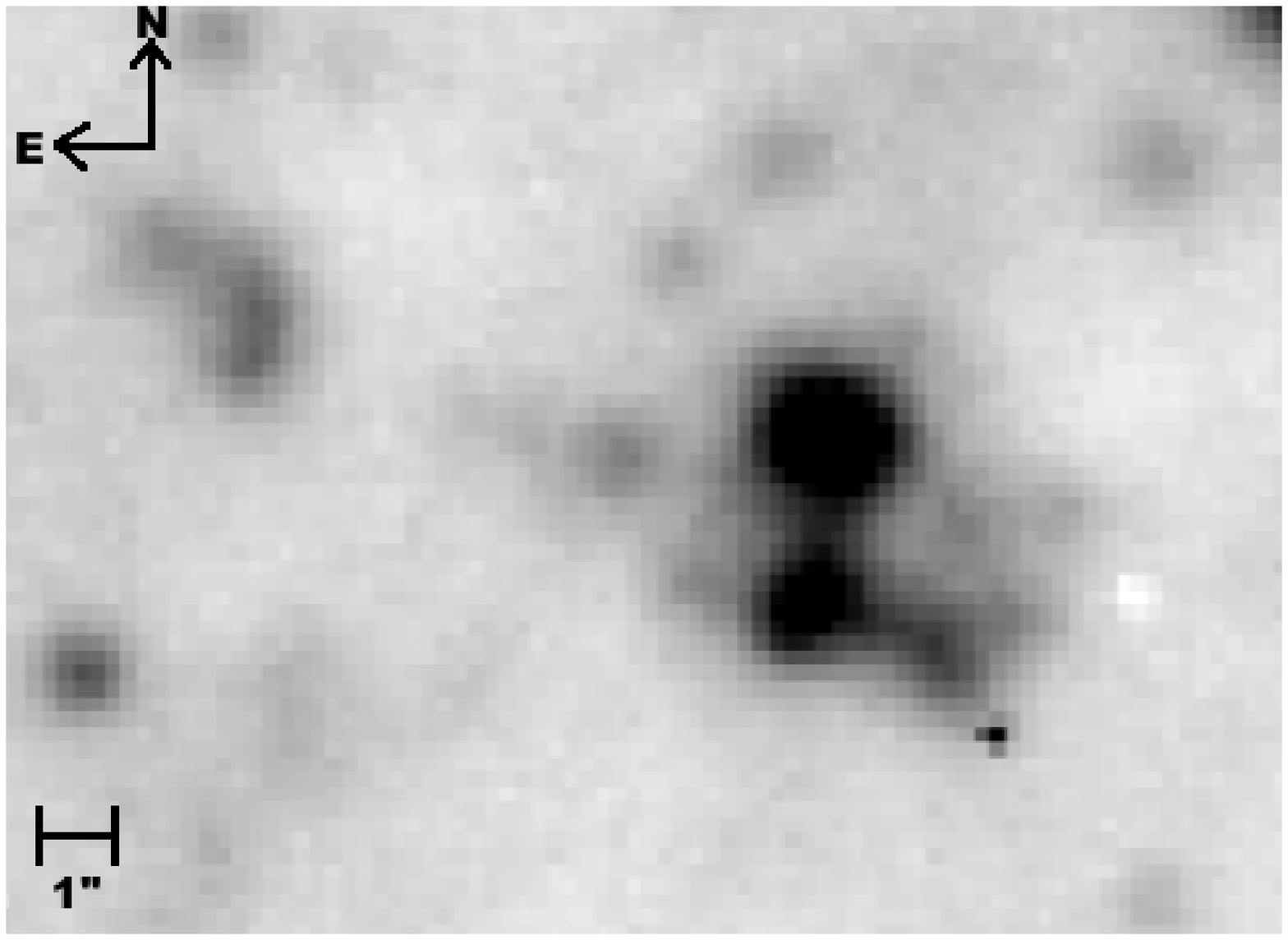}
                    \label{vlt-zoom}}

\subfigure[]{
  \includegraphics[width=0.9\columnwidth]{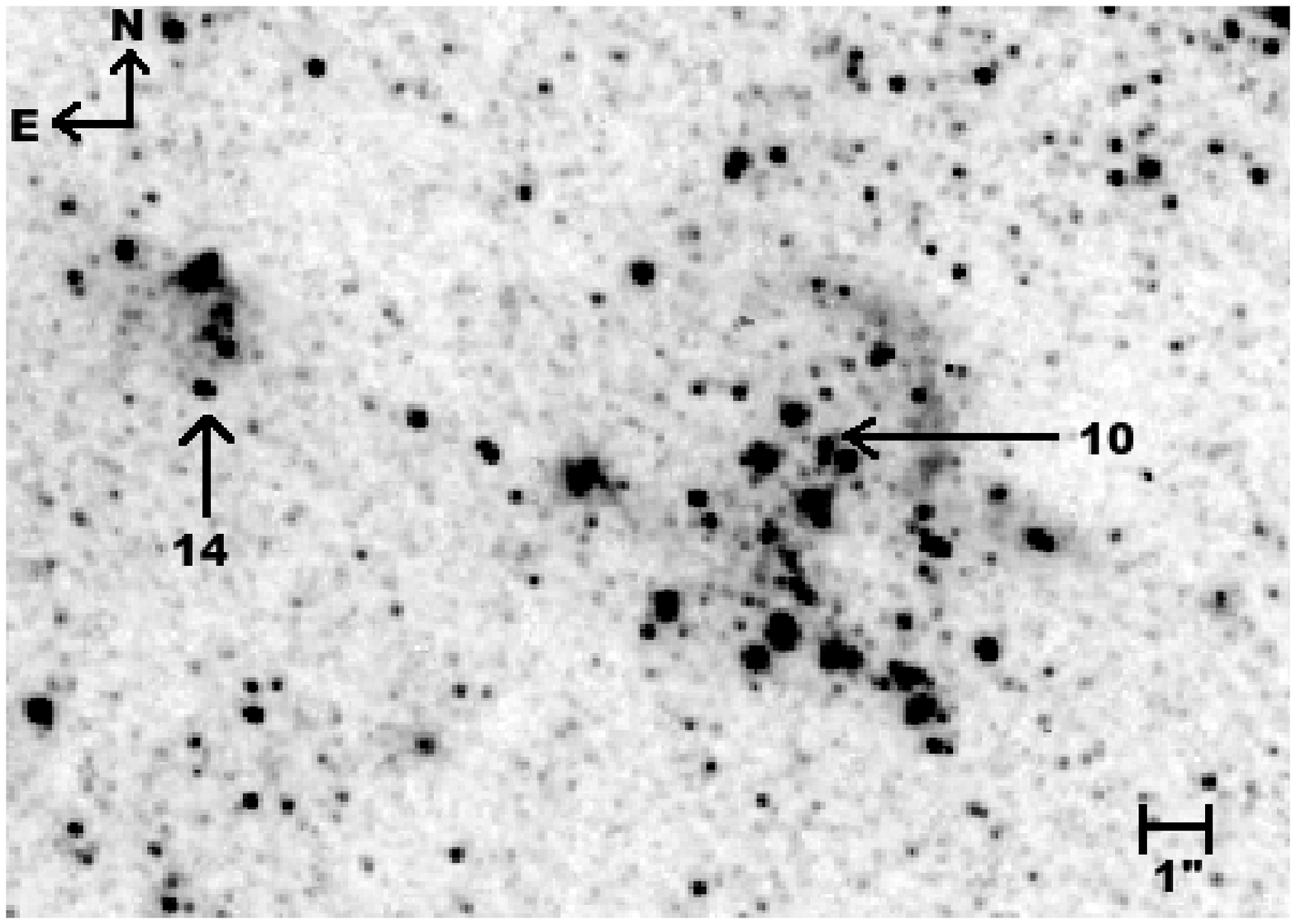}
                 \label{hst-zoom} }

\caption{Comparison of (a) ground-based $\lambda$~4684 VLT/FORS1
  imaging with (b) space-based HST/ACS F555W imaging. The image is
  $\sim$\,15\,$\times$ 10$\arcsec$ (300 $\times$ 200pc). From archival
  HST images we are able to locate the WR source more accurately than
  using ground-based data.}
  \label{spatialresolution}
   \end{figure}

33 of our 39 spectra showed WR features for which we have
obtained photometry from ground-based VLT images. However, crowding
and the spatial resolution of He\,{\sc ii} ground-based images cause
these sources to be contaminated by stars along similar lines of
sight, or stars within the same cluster. For NGC~7793 at a distance of
3.91\,Mpc \citep{Karachentsev2003} the spatial resolution of our
ground-based He\,{\sc ii} images is 1.3$\arcsec$ corresponding to a
physical scale of $\sim$25pc compared to 0.1$\arcsec$ resolution, or
$\sim$2pc, with HST. The dramatic improvement of space-based images is
demonstrated in Figure \ref{spatialresolution}. We can take advantage
of archival HST imaging of NGC~7793 (Section 2.2) to more accurately
locate the WR source and determine a more robust
magnitude. Unfortunately, not all of NGC~7793 has been observed with
ACS/WFC using the F555W filter. 10 of our candidates lie within the
ACS pointing (recall Figure~\ref{fov}).

We are able to spatially resolve the WR source for 6 of the 10 cases
using offsets from nearby point sources to the peak of the He\,{\sc
  ii} emission identified from the archival VLT images. Two examples
are shown in Figure~\ref{spatialresolution}. Table~\ref{hstmags} shows
that the measured HST magnitudes for 5 WR soruces are 0.4-1.4
magnitudes fainter, arising from severe crowding in the VLT
images. Source \#36 was not identified by \textsc{daophot} in the VLT
V-band image, although a HST magnitude can be determined. Sources \#8
and \#34 can only be partially resolved from the compact host cluster
in the HST image, making little difference to the observed
magnitude. The final two WR sources (\#24 and \#26) are located in
dense clusters, so despite the superior spatial resolution of HST, we
cannot identify the individual WR source. If HST/ACS F555W imaging
were available for all 74 sources in NGC~7793, magnitudes would typically
be 1$\pm$0.5 mag fainter than those quoted in Table \ref{sources}.

\begin{table}
\caption{HST F555W magnitudes compared with VLT V-band magnitudes for
  WR sources in NGC~7793. The HST magnitudes are fainter due to an
  increased spatial resolution so the multiple objects are no longer
  included in the aperture. The RA and Dec values correspond to the
  HST position of the WR source.}
\begin{center}
\begin{tabular}{c@{\hspace{2mm}}c@{\hspace{2mm}}c@{\hspace{2mm}}c@{\hspace{2mm}}c@{\hspace{2mm}}c} 
\hline
ID & RA  & Dec  & m$_V$ & m$_{F555W}$ & m$_{F555W}$-m$_V$ \\
   & J2000 & J2000 & mag & mag & mag  \\
\hline
\# 4  & 23:57:37.133 & -32:35:01.33 & 22.48  & 23.91 & 1.43 \\
\# 8  & 23:57:39.550 & -32:37:24.32 & 20.97  & 20.78 & --0.19* \\
\# 10 & 23:57:40.809 & -32:35:36.21 & 20.18  & 21.43 & 1.25 \\
\# 14 & 23:57:41.421 & -32:35:35.49 & 21.27  & 22.28 & 1.01 \\
\# 20 & 23:57:45.325 & -32:36:05.17 & 22.50  & 22.90 & 0.4 \\
\# 24 & 23:57:46.647 & -32:36:00.28 & 21.99  & --    & --$\dagger$  \\
\# 25 & 23:57:46.787 & -32:34:05.80 & 21.81  & 22.60 & 0.79 \\
\# 26 & 23:57:47.049 & -32:35:48.11 & 22.49  & --    & -- $\dagger$ \\
\# 34 & 23:57:48.830 & -32:34:53.10 & 18.07  & 18.09 & 0.02* \\
\# 36 & 23:57:49.037 & -32:34:57.41 & --     & 23.10 & -- \\
\hline
\multicolumn{6}{l}{* WR source only partially resolved from host cluster}\\
\multicolumn{6}{l}{$\dagger$ WR source is not resolved from host cluster} \\

\end{tabular}
\label{hstmags}
\end{center}
\end{table}

\begin{figure}
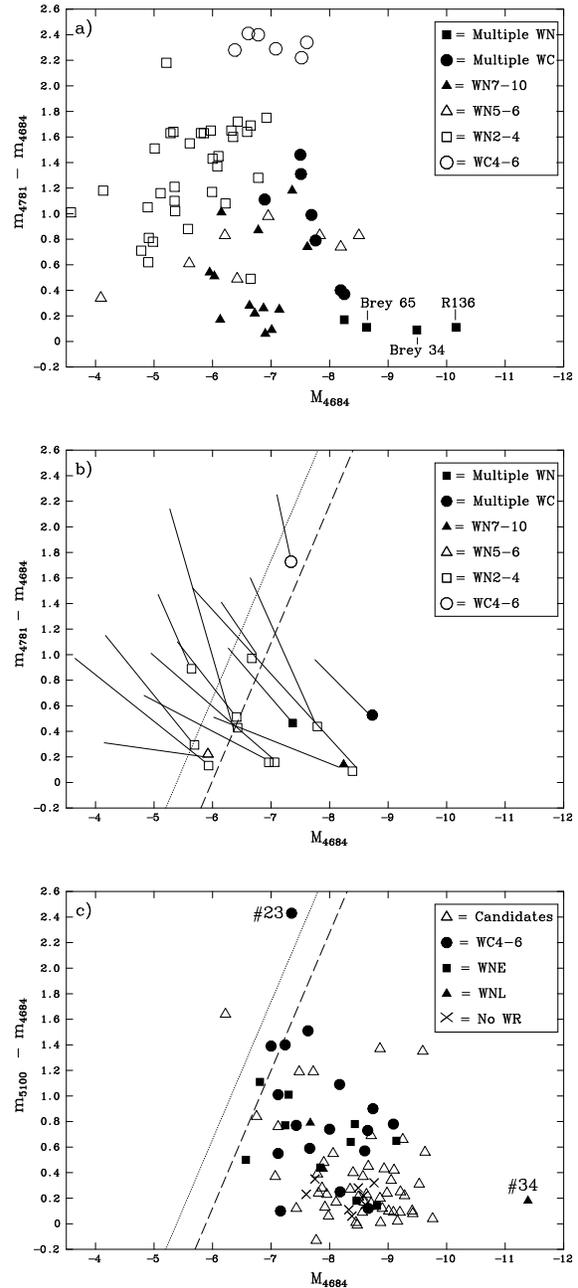

   \centering

\subfigure{
  \includegraphics[width=0.65\columnwidth, angle=-90]{lmc_wr_mag.ps}
                    \label{lmc}}

\subfigure{
  \includegraphics[width=0.65\columnwidth, angle=-90]{lmc_at_4Mpc_true.ps}
                    \label{lmc_4Mpc}}

\subfigure{
   \includegraphics[width=0.65\columnwidth, angle=-90]{ngc7793_wr_mag.ps}
                    \label{wr_ngc7793}}

\caption{Comparison of $\lambda_c$=4684 excess emission versus
  M$_{4684}$ for (a) WR stars in the LMC, (b) a subset of the LMC WR
  population at a distance of 4\,Mpc accounting for the contribution
  of sources within 1 arcsecond (20\,pc) and (c) NGC~7793. The solid
  lines in (b) show the photometric shift arising from an increased
  aperture size. The diagonal lines in (b) and (c) indicate our 100\%
  (dashed) and 50\% (dotted) detection limits. Each sample is split
  into different subtypes}
\label{dilution}
\end{figure}

\subsection{Completeness compared to the LMC}

We can investigate the completeness of our NGC~7793 survey by
degrading observations of a nearby galaxy, that has been completely
surveyed for WR stars, to the required resolution. We use ground-based
DSS \footnotemark ~images of the LMC, which lies at a distance of
$\sim$50 kpc \citep{Gibson2000}, to assess the role played by nearby
sources on the detection limit for WR stars for a sample of the LMC WR
stars listed in \citet{Breysacher1999}.

\footnotetext{The Digitized Sky Surveys were produced at the Space
  Telescope Science Institute under U.S. Government grant NAG
  W-2166. The images of these surveys are based on photographic data
  obtained using the Oschin Schmidt Telescope on Palomar Mountain and
  the UK Schmidt Telescope. The plates were processed into the present
  compressed digital form with the permission of these institutions.}

Figure \ref{lmc} shows the distribution of M$_{4684}$ and
m$_{4781}$-m$_{4684}$ for WR stars in the LMC. Typical broad-band
magnitudes of single LMC WR stars are M$_{V}$ = --3 to --7 mag.
 We have obtained synthetic narrow-band magnitudes for LMC WR stars
from spectrophotometric observations of \citet{Crowther2006} and
\citet{Torres-Dodgen&Massey1988}. We find m$_{4781}$-m$_{4684}$ =
0.1--1.8 mag for WN stars and 0.4--2.4 mag for WC stars. 
WR sources exceeding M$_{4684}$\,=\,--9 mag are clusters
containing both O stars and WR stars. For example R136 (Brey 82) is
the youngest, brightest cluster in the LMC \citep{Massey&Hunter1998},
Brey 34 is a WN star with a B supergiant companion \citep{Dopita1994},
whilst Brey 65 is a star cluster hosting a WN star
\citep{Walborn1999}. The He\,{\sc ii} emission from the WR stars
within these clusters/binaries has been severely diluted by the
companion stars, hence a relatively small excess is detected. Nevertheless,
the majority of the WR stars detected in the LMC can be resolved into
single stars.

However, if the LMC were located at a similar distance to NGC 7793,
then a higher percentage of the WR sources would be blended with the
surrounding stars if they were observed at the same spatial resolution
as our archival VLT/FORS imaging. Would the WR emission still be
observed, or would it be diluted by surrounding stars to the point
that it was no longer detectable?

To address this question we consider the location of a representative
sample of 15 WR stars in the LMC to investigate the effect of a
degraded resolution. 1 arcsecond corresponds to a spatial scale of
$\sim$\,0.25pc at the distance of the LMC versus $\sim$\,20pc for
NGC~7793. We have used DSS images to determine how the photometric
properties would alter if the surrounding stars were combined with the
WR star.

Figure \ref{lmc_4Mpc} shows the resulting photometry of the He\,{\sc
  ii} excess and absolute magnitude of this subset of WR stars in the
LMC at a distance of 4\,Mpc. The solid lines indicate the shift of the
WR stars to the lower right of the figure. Taking into account our
formal completeness limit of M$_{4684}\sim$--5.8 mag for NGC 7793, and
assuming that we include only sources with
M$_{4781}$-M$_{4684}$$\geq$0.1 mag we would detect at least 80\% of
the WR stars in the LMC in our VLT imaging survey. Moreover, our 50\%
detection limit includes all but one of the LMC WR stars in our
sample. Figure \ref{lmc_4Mpc} also shows that there is a bias towards
WC stars \citep{Massey&Johnson1998}, such that we would most likely
have detected the overwhelmingly majority of the LMC WC stars at 4~Mpc
(See Section 8). Figure \ref{wr_ngc7793} shows the photometric
absolute magnitude and He\,{\sc ii} excess emission for the WR sources
in NGC~7793, with the diagonal line representing the 50\% and 100\%
completeness limits (recall Section 2.3).

\section{Giant H\,{\sc ii} regions}\label{GHR}
NGC 7793 contains 132 catalogued H\,{\sc ii} regions
\citep{Davoust1980}. Using appropriate aperture radii (r$_{ap}$), we
have measured the net H$\alpha$ flux for the brightest H\,{\sc ii}
regions, limiting our sample to those which exceed Q$_0 >$10$^{50}$
photon s$^{-1}$, formally defined as giant H\,{\sc ii} regions (GHR)
in \citet{Conti2008}. From comparison with the CTIO images used in
\citet{Kennicutt&Lee2008} there are 4 additional giant H\,{\sc ii}
regions in NGC 7793 beyond the field of view of our VLT imaging
(recall Figure \ref{fov}). Two of these are listed in the
\citet{Davoust1980} catalogue as D47 and D132, while the other two
have not been discussed. We add the location of the two other four
GHRs to Table~\ref{HIIregions} as GHR A, B, C and D.

H$\alpha$ fluxes estimated from observations using the H$\alpha$
narrow-band filter are contaminated by the [N\,{\sc
    ii}]\,$\lambda$6583 emission, and to a lesser degree by [N\,{\sc
    ii}]\,$\lambda$6548. To correct for this we have determined the
[N\,{\sc ii}]\,$\lambda$6583 contribution from our spectroscopic data,
or adopt an average value of I([N\,{\sc
    ii}]\,$\lambda$6583)/I(H$\alpha$)=0.19 otherwise. 97\% of our
I([N\,{\sc ii}]\,$\lambda$6583)/I(H$\alpha$) values are within 1$\sigma$
(0.2) of the mean value. \citet{Webster1983} obtain a mean value
I[N\,{\sc ii}]\,$\lambda$6583/I(H$\alpha$) = 0.23 from spectroscopy of
H\,{\sc ii} regions, while \citet{McCall1985} find a higher mean value
of I([N\,{\sc ii}]\,$\lambda$6583)/I(H$\alpha$)=0.35 from the average of
only 3 H\,{\sc ii} regions. After subtracting the continuum, observed
fluxes were extinction corrected. Table \ref{HIIregions} lists values
of E(B-V) and I([N\,{\sc ii}]\,$\lambda$6583)/I(H$\alpha$) for each source.

We have calculated a H$\alpha$ luminosity of 3.30$\times$10$^{40}$erg
s$^{-1}$ or a Star Formation Rate (SFR)$\sim$0.26 M$_{\odot}$yr$^{-1}$
(using equation 2 in \citealt{Kennicutt1998}) for the region surveyed
with FORS1 (recall Figure~\ref{fov}).  A more accurate value for the
whole galaxy can be made by applying our mean values of E(B-V) and
I([N\,{\sc ii}]\,$\lambda$6583)/I(H$\alpha$) to the observed
(H$\alpha$+[N\,{\sc ii}]\,$\lambda$6583) flux of
\citet{Kennicutt&Lee2008}. The revised SFR =
0.45~M$_{\odot}$yr$^{-1}$, 50\% larger than their
SFR$\sim$0.30~M$_{\odot}$yr$^{-1}$, suggesting that emission beyond
our H$\alpha$ survey contributes $\sim$40\% of the total. We note that
young H\,{\sc ii} regions could be visibly obscured, increasing the
global SFR. However, NGC~7793 is included in the SINGS survey
\citep{Kennicutt2003}, for which \citet{Prescott2007} concluded that
NGC~7793 has no highly obscured bright H\,{\sc ii} regions. Therefore,
the H$\alpha$--derived SFR should reflect the true value.

\begin{table*}
\caption{ Catalogue of Giant H\,{\sc ii} Regions (GHRs) in
  NGC~7793. All luminosities are based on a distance of 3.91~Mpc
  \citep{Karachentsev2003}. F(H$\alpha$) is the continuum subtracted,
        [N\,{\sc ii}]\,$\lambda$6583 corrected observed flux.  We note
        that there are $\sim$4 additional GHII regions within the
        outer regions of NGC~7793 beyond our VLT/FORS1 H$\alpha$
        surveyed region, two of which are listed in the
        \citet{Davoust1980} catalogue. The final two rows relate to
        our H$\alpha$ survey region, and global properties for
        NGC~7793 updated from \citet{Kennicutt&Lee2008}.  }
\begin{tabular}{c@{\hspace{1.5mm}}c@{\hspace{1.5mm}}c@{\hspace{1.5mm}}c@{\hspace{1.5mm}}c@{\hspace{1.5mm}}c
@{\hspace{1.5mm}}c@{\hspace{1.5mm}}c@{\hspace{1.5mm}}c@{\hspace{1.5mm}}c@{\hspace{1.5mm}}c@{\hspace{1.5mm}}c}
\hline
ID & RA & Dec & r$_{ap}^{1}$ & E(B-V) & I[N\,{\sc ii}]/I(H$\alpha$)  &  F(H$\alpha$) & L(H$\alpha$) & log 
Q$_{0}$ & N(O7V) &  HII & WR \\
       &J2000    &J2000     & arcsec & mag    &                & erg s$^{-1}$ cm$^{-2}$ & erg s$^{-1}$ & s$^{-1}$ 
& &    region  &               \\
\hline
GHR $\#1$ & 23:57:41.166 & -32:35:50.89 & 6.25 & 0.179 & 0.19 & 1.80$\times$10$^{-13}$ & 5.00$\times$10$^{38}$ & 
50.57 
& 37 & D13 & 12ab \\
GHR $\#2$ & 23:57:41.237 & -32:34:51.45 & 3.75 & 0.179 & 0.16 & 1.40$\times$10$^{-13}$ & 3.89$\times$10$^{38}$ & 
50.46 
& 29 & D14 & 11a--c \\
GHR $\#3$ & 23:57:48.889 & -32:34:53.13 & 3.00 & 0.319 & 0.26 & 8.24$\times$10$^{-14}$ & 3.17$\times$10$^{38}$ & 
50.37 & 23 & -- & 34$^{3}$ \\
GHR $\#4$ & 23:57:54.250 & -32:33:59.78 & 4.75 & 0.242 & 0.18 & 9.60$\times$10$^{-14}$ & 3.09$\times$10$^{38}$ & 
50.36 
& 23 & D95 & 46$^{3}$ \\
GHR $\#5$ & 23:57:43.271 & -32:35:49.90 & 4.75 & 0.179 & 0.19 & 1.03$\times$10$^{-13}$ & 2.86$\times$10$^{38}$ & 
50.33 
& 21 & D20 & 17 \\
GHR $\#6$ & 23:57:46.711 & -32:36:06.69 & 3.50 & 0.179 & 0.19 & 9.27$\times$10$^{-14}$ & 2.58$\times$10$^{38}$ & 
50.28 
& 19 & D36 & -- \\
GHR $\#7$ & 23:57:57.041 & -32:36:08.62 & 4.50 & 0.179 & 0.19 & 8.55$\times$10$^{-14}$ & 2.38$\times$10$^{38}$ & 
50.25 
& 18 & D110 & 53 \\
GHR $\#8$ & 23:57:56.968 & -32:33:48.35 & 4.25 & 0.179 & 0.19 & 8.08$\times$10$^{-14}$ & 2.25$\times$10$^{38}$ & 
50.22 
& 17 & D111 & 52ab \\
GHR $\#9$ & 23:57:50.798 & -32:34:17.55 & 4.25 & 0.179 & 0.19 & 7.72$\times$10$^{-14}$ & 2.15$\times$10$^{38}$ & 
50.20 
& 16 & D73 & -- \\
GHR $\#10$& 23:57:41.416 & -32:35:34.74 & 2.75 & 0.298 & 0.07 & 5.15$\times$10$^{-14}$ & 1.89$\times$10$^{38}$ & 
50.15 
& 14 & D15 & 14$^{3}$ \\
GHR $\#11$& 23:57:48.209 & -32:36:15.03 & 2.50 & 0.179 & 0.19 & 6.64$\times$10$^{-14}$ & 1.85$\times$10$^{38}$ & 
50.14 
& 14 & D49 & -- \\
GHR $\#12$& 23:57:51.194 & -32:36:48.75 & 2.75 & 0.179 & 0.19 & 5.76$\times$10$^{-14}$ & 1.60$\times$10$^{38}$ & 
50.07 
& 12 & D74 & -- \\
GHR $\#13$& 23:57:44.744 & -32:34:24.99 & 2.75 & 0.179 & 0.19 & 5.68$\times$10$^{-14}$ & 1.58$\times$10$^{38}$ & 
50.07 
& 12 & D27 & -- \\
GHR $\#14$& 23:57:44.494 & -32:35:52.03 & 3.00 & 0.119 & 0.12 & 6.50$\times$10$^{-14}$ & 1.57$\times$10$^{38}$ & 
50.07 
& 12 & D25 & 22a--c \\
\smallskip
GHR $\#15$& 23:57:48.391 & -32:34:34.25 & 3.00 & 0.179 & 0.19 & 4.93$\times$10$^{-14}$ & 1.37$\times$10$^{38}$ & 
50.01 
& 10 & D55 & 30ab \\
GHR A & 23:57:48.000 & -32:33:04.10 & & & & & & & & D47 & -- \\
GHR B & 23:58:06.730 & -32:34:58.00 & & & & & & & & D132 & 73ab \\
GHR C & 23:57:59.967 & -32:33:23.84 & & & & & & & & & --\\
GHR D & 23:58:08.807 & -32:36:47.68 & & & & & & & & & -- \\

\hline
\multicolumn{3}{l}{H$\alpha$ survey (This study)}             & 205.7& 0.179 & 0.19 & 1.18$\times$10$^{-11}$ & 
3.30$\times$10$^{40}$ & 52.39 
& 
2440 & \multicolumn{2}{c}{SFR = 0.26 M$_{\odot}$yr$^{-1}$} \\
\multicolumn{3}{l}{H$\alpha$ survey (\citealt{Kennicutt&Lee2008})} & &  0.179 & 0.19 & 2.04$\times$10$^{-11}$ & 
5.65$\times$10$^{40}$ & 52.62 & 4180 & 
\multicolumn{2}{c}{SFR = 0.45 M$_{\odot}$yr$^{-1}$} \\
\hline
\multicolumn{4}{l}{$^{1}$ \footnotesize{Aperture radius}} \\

\end{tabular}
\label{HIIregions}
\end{table*}

\section{A background quasar: Q2358-32}

\begin{figure}
  \centering
   \includegraphics[width=0.7\columnwidth,angle=-90]{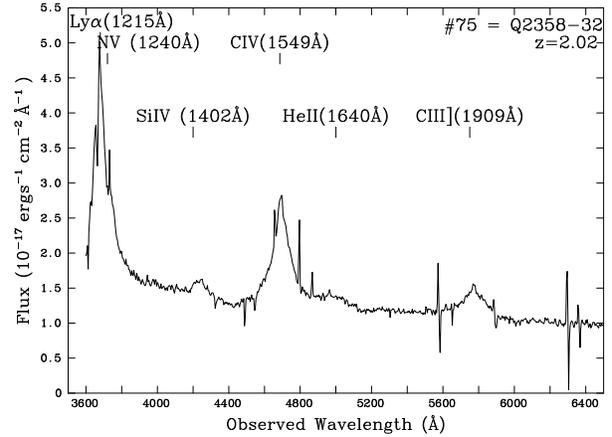}
     \caption{FORS1 spectrum of the background quasar Q2358-32 which
       lies at $z$=2.02. This source was picked up in our WR
       survey because C\,{\sc iv} $\lambda$1548-51 was
       redshifted into the $\lambda_c$=4684$\AA$ narrow-band filter.}
                   \label{quasar}
   \end{figure}

Follow-up MOS spectroscopy of one candidate emission line source
  displaying a large excess of m$_{4684}$-m$_{5100}$\,=\,--0.62 mag failed
  to match that expected for a WR star. This source, for which
  m$_V$=20.79 mag, was revealed instead to be a background quasar at
  $z$\,$\sim$2.02 in which C\,{\sc iv} $\lambda$1548--51 has been
  redshifted into the $\lambda$4684 narrow-band filter. The quasar
  spectrum is presented in Fig.~\ref{quasar}, which we name Q2358-32.

NGC 7793 has been observed with the X-ray telescope aboard ROSAT
\citep{ReadPietsch99}. From this survey the published coordinates of
X-ray point source P11 lies $\sim$20 arcsec from Q2358-32. Although this
source has a formal positional error of only 13 arcsec, P11 could
plausibly arise from Q2358-32 from inspection of figure~1 in
\citet{ReadPietsch99}.

A number of quasars located behind nearby spiral galaxies are
known, including a $z$\,=\,2.55 quasar towards NGC~1365 by \citet{Bresolin2005}.
These have previously been used to investigate the nature of
the interstellar medium of the spiral galaxy using the Ca\,{\sc ii} H
and K lines - see \citet{Pettini&Boksenberg1985} and
\citet{Hintzen1990}. Unfortunately, in the case of Q2358-32 these
lines fall on the redshifted Lyman $\alpha$ QSO emission line, hence
we are unable to calculate the column density of the absorbing gas in
this instance \citep{Bowen1991}.  

\section{Global WR population of NGC 7793 with Comparison to NGC 300.}

\subsection{Global WR population of NGC 7793}

 We confirm the presence of 27 WN and 25 WC stars in NGC 7793 from
  spectroscopic observations (Table~\ref{sources}). However to
  estimate the global number of WR stars in NGC 7793 we may use (a)
  our narrow-band photometry of remaining candidates and (b) our
  comparison with the LMC WR population.

We photometrically classify remaining WR candidates in NGC 7793 as
follows, on the basis of the distribution of LMC WR stars in
Figure~\ref{lmc_4Mpc}. We shall assume that sources with
m$_{5100}$-m$_{4684}$$>$+1 mag are WC subtypes, those for which
+0.2$\leq$m$_{5100}$-m$_{4684} \leq$+1 mag are WN stars, with the
remainder either foreground or weak emission-line sources. In reality
there will be instances of WC, WN and spurious candidates in different
magnitude bins, but outliers should be relatively few in number. In
addition, we uniformly assume that photometric sources host a solitary
WR star. A total of 27 WN and 8 WC candidates are added to the census
in this way, listed in parentheses in Table~\ref{sources}, suggesting
54 WN and 33 WC stars in total.

Finally, we need to consider the completeness limit of our imaging survey. We estimate
that 90\% of the WC stars in the region of NGC~7793 surveyed would have been detected,
plus 80\% of the WN stars, on the basis that their intrinsic properties are similar 
to those in the LMC. Adjusting the WR populations for this effect, we arrive at 
a total of $\sim$68 WN and $\sim$37 WC stars, with N(WC)/N(WN)$\sim$0.5 and N(WR)=105 
in total. Our
spectroscopic
survey has therefore recovered 67\% of the estimated WC population
and 40\% of the estimated WN population. Using the revised global star formation rate 
from Sect.~\ref{GHR} we obtain N(WR)/N(O7\,V)$\sim$0.025. 

\subsection{Comparison between WR populations of NGC 7793 and NGC 300}

Wolf-Rayet surveys have been conducted within both Local Group galaxies 
\citep{Massey&Johnson1998} and beyond. Of those non-Local Group 
spiral galaxies forming the basis of our supernova progenitor survey, M83
is metal-rich \citep{Hadfield2005}, NGC 1313 is metal-poor \citep{Hadfield2007},
leaving the central region of NGC 300 surveyed by \citet{Schild2003}
as the ideal comparison galaxy to the present survey of NGC 7793.

NGC 300 is another Sculptor group SA(s)d spiral galaxy, albeit a factor of two
closer \citep{Gieren2005}. Basic global properties of NGC 300 and NGC 7793
are provided in Table~\ref{ngc300}, indicating similar physical sizes and metallicity
gradients, although NGC 7793 possesses a factor of four higher star formation rate
(Sect.~\ref{GHR}). We also provide a comparison between the WR census
of the central region of NGC 300 and the present study. 30 of the 58 WR candidates 
identified by \citet{Schild2003} have been spectroscopically confirmed by references
therein or \citet{Crowther&Carpano2007}. In addition, \citet{Crowther2010} have
recently obtained additional MOS spectroscopy of NGC~300 from which 5 additional
WR stars are revealed (\#7, \#10, \#21, \#37 and \#54), to which we add the
confirmation of a very late WN star discovered by \citet{Bresolin2009},
bringing the total to 20 WN and 17 WC stars, including a couple of composite
WN+WC systems (\#11, \#37). We set a slightly lower $\lambda$4684 excess threshold of 
 m$_{4781}$-m$_{4684}$ = +0.15 mag  
for  the remaining photometric candidates in NGC 300 as a result of an improved photometric precision
and slightly different continuum filter. Six additional photometric WN stars follow.
These bring the total WR census of the inner disk of NGC 300 to N(WR)$\sim$43, comprising
26 WN and 17 WC stars, or N(WC)/N(WN)$\sim$0.65.

Regarding the O star content of
the central region of NGC~300, \citet{Crowther&Carpano2007} estimated that the
WR survey region included $\sim$78\% of the 0.12 $M_{\odot}$ yr$^{-1}$ 
global star formation rate of NGC~300 \citep{Kennicutt&Lee2008}. Therefore,
N(WR)/N(O7V)$\sim$0.05, a factor of two higher than NGC~7793. Indeed the surface
density of WR  stars in the central region of NGC~300 is also a factor of two higher
than the entire disk of NGC~7793. This difference may originate in part
from the slightly higher metallicity of the inner
region of NGC~300, namely $\log$ (O/H) + 12 $\sim$ 8.5, versus an average
value of $\sim$8.4 for NGC~7793.

\begin{table}
\caption{Global properties of NGC~7793 with respect to NGC~300, plus a detailed
census of their massive star content within the regions surveyed by \citet{Schild2003}
and this study. The values in  parentheses include adjustments for sources lacking
spectroscopy plus completeness issues for NGC~7793.}
\begin{tabular}{c|cc}
Name                                          & NGC 300               & NGC 7793    \\
\hline
Hubble Type                              & SA(s)d                    & SA(s)d \\
Distance (Mpc)                            & 1.88$^a$               & 3.91$^h$\\
$R_{25}$ (arcmin)                          & 10.9$^b$               & 4.65$^{b}$ \\
R$_{25}$ (kpc)                                  & 5.3$^b$                & 5.3$^b$ \\
SFR (M$_{\odot}$yr$^{-1}$)                 & 0.12$^d$                & 0.45$^{d,i}$ \\
log(O/H) + 12 (centre)                    & 8.57$^e$                & 8.61$^i$ \\
Gradient$^*$    (dex kpc$^{-1}$)           & --0.08$^f$             & --0.07$^i$ \\
\hline
Survey region (kpc)                     & 3.7 $\times$ 3.7$^c$    & 9.3 $\times$ 7.3$^i$ \\
N(O7V)                                    & 870$^{i}$               & 4200$^{d,i}$ \\
N(O)                                      & 1300$^{i}$               & 6250$^{d,i}$ \\
$\log$(O/H)+12 (mean)                     & 8.5$^{e}$               & 8.4$^{i}$ \\
WR candidates                             & 58$^{c f}$               & 74$^i$  \\
N(WN)                                     & 20 (26)$^{c f g j}$       & 27 (68)$^{i}$ \\
N(WC)                                     & 17 (17)$^{c f g j}$       & 25 (37)$^{i}$ \\
N(WC)/N(WN)                               & 0.85 (0.65)$^{i}$         & 0.93 (0.55)$^{i}$ \\
N(WR)/N(O)                              & $\geq$0.028 (0.033)$^i$     & $\geq$0.008 (0.017)$^i$ \\
Surface density$^*$ (WR kpc$^{-2}$)        & 2.7 (3.1)$^i$                 & 0.77 (1.6)$^i$ \\
\hline
\multicolumn{3}{l}{$^a$ \citet{Gieren2005}, $^b$ \citet{deVaucouleurs1991}, $^c$ \citet{Schild2003}} \\
\multicolumn{3}{l}{$^d$ \citet{Kennicutt&Lee2008}, $^e$ \citet{Urbaneja2005}, $^f$ \citet{Bresolin2009}} \\
\multicolumn{3}{l}{$^g$ \citet{Crowther&Carpano2007}, $^h$ \citet{Karachentsev2003}, $^i$ This work}\\
\multicolumn{3}{l}{$^j$ \citet{Crowther2010}} \\

\label{ngc300}
\end{tabular}
\end{table}

\subsection{Comparison with evolutionary models}

\begin{figure}
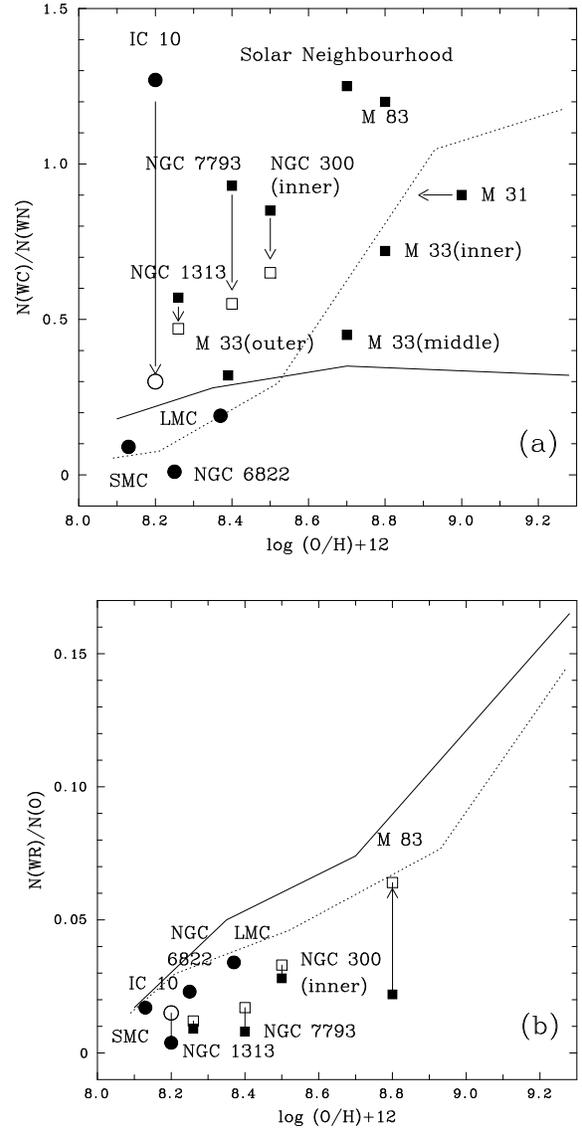

   \centering
\subfigure{
   \includegraphics[width=0.9\columnwidth]{WC_WN.eps}
\label{wc_wn}}

\subfigure{
  \includegraphics[width=0.9\columnwidth]{WR_O.eps}
\label{wr_o}}

    \caption{Comparison between (a) N(WC)/N(WN) and (b) N(WR)/N(O)
      ratios and the oxygen content of nearby spiral (squares) and
      irregular (circles) galaxies, together with evolutionary
      predictions from \citet{Meynet&Maeder2005} and
      \citet{EldridgeVink06}.  Spectroscopic (filled symbols) and
      photometric (open symbols) results are shown for NGC 7793, NGC
      300, NGC 1313 and IC 10 from the present study,
      \citet{Hadfield2007}, \citet{Massey&Holmes2002} and
      \citet{Crowther2003}.}
              
\label{evolution}     
   \end{figure}

Observations of WR stars in the Local Group have revealed a strong
correlation between the ratio of WC to WN stars and WR to O stars
as a function of metallicity \citep{Massey1996}. In
Figure~\ref{wc_wn} we provide an updated comparison of N(WC)/N(WN)
including spectroscopic and photometry results from our galaxy surveys to date, 
plus the photometric results of \citet{Massey&Holmes2002}. Model
predictions are shown from initially rotating (300 km\,s$^{-1}$) single star models 
of  \citet{Meynet&Maeder2005} plus single and binary models incorporating
metallicity-dependent WR winds from \citet{EldridgeVink06}. The latter
provides a reasonable match to empirical results, although highlights differences
between the metal-poor 
dwarf spiral galaxies (including NGC 7793) and dwarf irregular galaxies (including
the LMC).

In order to compare the ratio of WR to O stars with evolutionary predictions,
it is necessary to first calculate the number of equivalent O7V stars from 
nebular H$\alpha$ luminosities, and then attempt to convert this quantity into 
actual O stars. For the LMC and SMC we utilise nebular properties from 
\citet{Kennicutt1995}, with H$\alpha$ luminosities from \citet{Lee2009} used,
with the exceptions of IC~10 \citep{GildePaz2003}, NGC 300 and NGC 7793 (recall
Section~8.2). These are converted into the number of equivalent O7V stars using
equation~2 from \citet{Kennicutt1998} and $Q_{0}$ = 10$^{49}$ s$^{-1}$ for O7V stars.

For an instantaneous burst of star formation, N(O)/N(O7V)$\sim$0.7--2
for the first $\sim$4 Myr at LMC metallicity
\citep{Schaerer&Vacca1998}. For continuous star formation, as is the
case for galaxies in our sample, we have estimated the conversion
factors in two ways. There are $\sim$1000 O stars in the SMC according
to \citet{Evans2004}, suggesting N(O)/N(O7V)$\sim$2 from its global
H$\alpha$ derived star-formation rate. For the 30 Doradus region of
the LMC, the recent VLT-FLAMES Tarantula survey \citep{Evans2010}
reveals $\sim$400 O stars excluding the central R136 ionizing cluster
\citep{Crowther&Dessart1998}.  30 Doradus provides a total Lyman
continuum ionizing flux of $\sim 10^{52}$ s$^{-1}$, of which
$\sim$40\% is supplied by R136, suggesting N(O)/N(O7V)$\sim$0.7 from
30 Doradus (omitting R136).  Figure~\ref{wr_o} compares the WR to O
ratio for galaxies within our survey and results from the literature
\citep{Massey1996} applying a uniform correction of
N(O)/N(O7V)$\sim$1.5 for the whole sample. Once again, evolutionary
predictions from \citet{EldridgeVink06} and \citet{Meynet&Maeder2005}
are included, with reasonable agreement found for both, albeit
highlighting the relatively low WR to O ratio for NGC~7793.

\section{Summary}

We present the results of a VLT/FORS1 imaging and spectroscopic
  survey of the WR population of the Sculptor group spiral
  galaxy NGC~7793.
\begin{enumerate}
\item From archival narrow-band imaging, we identify 74 candidate
  emission line regions, of which 39 have been spectroscopically
  observed with the Multi Object Spectroscopy (MOS) mode of FORS1. Of
  these, 85\% of these sources exhibited WR features above a 3
  $\sigma$ level.  Additional slits were used to observe H\,{\sc ii}
  regions, enabling an estimate of the metallicity gradient of
  NGC~7793 using strong line calibrations, from which $\log$ (O/H) +
  12 = 8.61 $\pm$ 0.05 - (0.36 $\pm$ 0.10) r/R$_{\rm 25}$ was
  obtained. We have estimated WR populations using a
  calibration of line luminosities of Large Magellanic Cloud stars,
  revealing $\sim$27 WN and $\sim$25 WC stars for sources
  spectroscopically observed.
\item Photometric properties of the remaining candidates suggest an
  additional $\sim$27 WN and $\sim$8 WC stars. In addition, a
  comparison with LMC WR stars degraded to the spatial
  resolution achieved for NGC~7793 suggests that our imaging survey
  has identified $\sim$80\% of WN stars and $\sim$90\% for the WC
  subclass, from which a total of 68 WN and 37 WC stars are inferred
  within NGC~7793, i.e. N(WC)/N(WN)$\sim$0.5
\item Our H\,{\sc ii} region spectroscopy permits an updated star
  formation rate of 0.45 $M_{\odot}$ yr$^{-1}$ with respect to
  \citet{Kennicutt&Lee2008}, from which N(WR)/N(O)$\sim$0.017 is
  obtained, assuming N(O)/N(O7V)$\sim$1.5. 
A comparison between the WR census of NGC~7793 and survey
  of the central region of NGC 300 by \citet{Schild2003} is carried
  out. This reveals somewhat higher N(WR)/N(O) and N(WC)/N(WN) ratios
  in NGC~300, in part anticipated owing to its slightly higher mean
  metallicity.
\item NGC~7793 represents the fourth of ten star-forming spiral
  galaxies within 2--8 Mpc whose WR populations that we are
  surveying. Once completed, these will provide a database from which
  the nature of a future Type Ib/c supernova can be investigated.
\item Therefore, we have considered biases arising from differences
  between the intrinsic line strengths of WN and WC stars plus
  ground-based spatial resolution limitations. Therefore, we consider
  (a) the LMC WR population degraded to that if it was located
  at a distance of 4 Mpc; (b) differences to the apparent magnitude to
  the subset of NGC~7793 sources resulting from higher spatial
  resolution HST/ACS imaging. Upcoming narrow-band WFC3 Hubble Space
  Telescope imaging of the grand-design spiral galaxy M101
  (P.I. M.~Shara) will therefore provide the ideal dataset with which
  to assess the nature of its future core-collapse SN. This
  complements both our ground-based studies and the WFPC2 survey of
  the nearby late-type spiral NGC~2403 by \citet{Drissen99}.
\item Finally, we also report the fortuitous detection of a bright
  ($m_{\rm V}$ = 20.8 mag) background quasar Q2358-32 at $z \sim 2.02$
  resulting from C\,{\sc iv} $\lambda$1548-51 redshifted to the
  $\lambda$4684 passband.
\end{enumerate}

\section*{Acknowledgements}
JLB acknowledges financial support from STFC. Some of the data
presented in this paper were obtained from the Multimission Archive at
the Space Telescope Science Institute (MAST). STScI is operated by the
Association of Universities for Research in Astronomy, Inc., under
NASA contract NAS5-26555. Support for MAST for non-HST data is
provided by the NASA Office of Space Science via grant NAG5-7584 and
by other grants and contracts. Support for some of this work, part of
the Spitzer Space Telescope Legacy Science Programme, was provided by
NASA through contract 1224769 issued by the Jet Propulsion Laboratory,
California Institute of Technology under NASA contract 1407.

\begin{table*}
\setcounter{table}{2}
\caption{Observed, F$_{\lambda}$, and dereddened, I$_{\lambda}$
  nebular fluxes of H\,{\sc ii} regions in NGC~7793, relative to
  H$\beta$. The final row lists H$\beta$ fluxes in units of $\times
  10^{-15}$ erg s$^{-1}$cm$^{-2}$.}
\begin{tabular}{cccccccccccccc}
\hline

$\lambda$($\AA$) & ID & \multicolumn{2}{c}{10} & \multicolumn{2}{c}{11a+c} & \multicolumn{2}{c}{14} & \multicolumn{2}{c}{16}& \multicolumn{2}{c}{22a--b} & \multicolumn{2}{c}{23} \\

 &   & $F_{\lambda}$ & $I_{\lambda}$ &  $F_{\lambda}$ & $I_{\lambda}$ & $F_{\lambda}$ & $I_{\lambda}$ & $F_{\lambda}$ & $I_{\lambda}$ & $F_{\lambda}$ & $I_{\lambda}$  & $F_{\lambda}$ & $I_{\lambda}$  \\
\hline
3727  & [O\,{\sc ii}] &  217 & 251 & 145 & 145 & 90  & 116 & 189 & 236 & 166 & 184 & 471 & 526 \\
4343  & H$\gamma$     &  25  & 27  & 42  & 42  & 42  & 47  & 33  & 37  & 40  &  42 & 49  & 51  \\ 
4861  & H$\beta$      &  100 & 100 & 100 & 100 & 100 & 100 & 100 & 100 & 100 & 100 & 100 & 100 \\
4959  & [O\,{\sc iii}]&  126 & 125 & 60  & 60  & 76  & 74  & 147 & 143 & 53  & 53  & 108 & 107 \\
5007  & [O\,{\sc iii}]&  380 & 372 & 149 & 149 & 232 & 224 & 431 & 418 & 158 & 156 & 331 & 326 \\ 
6563  & H$\alpha$     &  345 & 289 & 231 & 231 & 400 & 291 & 386 & 292 & 327 & 288 & 331 & 288 \\
6583  & [N\,{\sc ii}] &  40  & 33  & 38  & 38  & 27  & 19  & 34  & 26  & 38  & 33  & 55  & 48  \\
6716  & [S\,{\sc ii}] &  27  & 22  & 17  & 17  & 7   & 5   & 45  & 34  & 18  & 16  & 44  & 38  \\
6731  & [S\,{\sc ii}] &  25  & 20  & 12  & 12   & 4   & 3   & 27  & 20  & 14  & 12  & 28  & 24  \\
\hline
4681  & H$\beta$  & 0.423 & 0.740 & 12.2 & 12.2 & 0.265 & 0.718 & 0.299 & 0.713 & 2.08 & 3.10 & 0.111 & 0.170 \\
\hline
\end{tabular}
\label{fluxes}
\begin{tabular}{cccccccccccccc}
\hline
$\lambda$($\AA$) & ID   &  \multicolumn{2}{c}{26} &\multicolumn{2}{c}{27} & \multicolumn{2}{c}{34} &\multicolumn{2}{c}{35} &\multicolumn{2}{c}{36} & \multicolumn{2}{c}{37} \\

 &    & $F_{\lambda}$ & $I_{\lambda}$ & $F_{\lambda}$ & $I_{\lambda}$ & $F_{\lambda}$ & $I_{\lambda}$ & $F_{\lambda}$ & $I_{\lambda}$  & $F_{\lambda}$ & $I_{\lambda}$ & $F_{\lambda}$ & $I_{\lambda}$ \\     
\hline
3727  & [O\,{\sc ii}] & 579 & 662 & 417 & 473 & 267 & 352 & 490 & 528 & 187 & 234 & 295 & 330 \\
4343  & H$\gamma$     & 37  & 40  & 45  &  47 & --  & --  & 21  & 21  & 41  & 46  & 48  & 50  \\ 
4861  & H$\beta$      & 100 & 100 & 100 & 100 & 100 & 100 & 100 & 100 & 100 & 100 & 100 & 100 \\
4959  & [O\,{\sc iii}]& 25  & 25  & 47  & 46  & 34  & 33  & 32  & 31  & 25  & 25  & 46  & 46  \\
5007  & [O\,{\sc iii}]& 70  & 68  & 145 & 142 & 100 & 96  & 95  & 94  & 76  & 74  & 139 & 137 \\ 
6563  & H$\alpha$     & 345 & 286 & 337 & 289 & 410 & 291 & 316 & 287 & 388 & 292 & 331 & 289 \\
6583  & [N\,{\sc ii}] & 158 & 131 & 74  & 64  & 103 & 74  & 88  & 80  & 91  & 68  & 48  & 42  \\
6716  & [S\,{\sc ii}] & 87  & 72  & 38  & 33  & 48  & 33  & 61  & 55  & 54  & 40  & 34  & 30  \\
6731  & [S\,{\sc ii}] & 56  & 46  & 32  & 27  & 32  & 22  & 52  & 47  & 32  & 23  & 22  & 19  \\
\hline
4681  & H$\beta$  & 0.067 & 0.118 & 0.166 & 0.270 & 1.29 & 3.76 & 0.141 & 0.189 & 0.466 & 1.14 & 1.02 & 1.57 \\
\hline
\end{tabular}
\label{fluxes}
\begin{tabular}{cccccccccccccc}
\hline
$\lambda$($\AA$) & ID &  \multicolumn{2}{c}{nucleus} & \multicolumn{2}{c}{39} & \multicolumn{2}{c}{42a--b} & \multicolumn{2}{c}{46}  & \multicolumn{2}{c}{47}& \multicolumn{2}{c}{49} \\
&    & $F_{\lambda}$ & $I_{\lambda}$ & $F_{\lambda}$ & $I_{\lambda}$ & $F_{\lambda}$ & $I_{\lambda}$ & $F_{\lambda}$ & $I_{\lambda}$  & $F_{\lambda}$ & $I_{\lambda}$ & $F_{\lambda}$ & $I_{\lambda}$ \\
\hline
3727  & [O\,{\sc ii}]  & 154 & 179 & --  & --  & 349 & 444 & 345 & 399 & 267 & 307 & 332 & 388 \\
4343  & H$\gamma$      & 29  & 31  & 31  & 34  & 45  & 51  & 41  & 44  & 46  & 49  & 50  & 54  \\ 
4861  & H$\beta$       & 100 & 100 & 100 & 100 & 100 & 100 & 100 & 100 & 100 & 100 & 100 & 100 \\
4959  & [O\,{\sc iii}] & 30  & 30  & 35  & 34  & 69  & 67  & 48  & 48  & 21  & 21  & 189 & 187 \\
5007  & [O\,{\sc iii}] & 84  & 82  & 107 & 104 & 211 & 180 & 145 & 142 & 71  & 70  & 558 & 546 \\ 
6563  & H$\alpha$      & 222 & 183 & 389 & 291 & 396 & 293 & 346 & 289 & 344 & 289 & 290 & 239 \\
6583  & [N\,{\sc ii}]  & 88  & 72  & 62  & 46  & 93  & 68  & 61  & 50  & 80  & 67  & 52  & 43  \\
6716  & [S\,{\sc ii}]  & 32  & 26  & 42  & 31  & 59  & 43  & 47  & 39  & 39  & 33  & 49  & 40  \\
6731  & [S\,{\sc ii}]  & 22  & 18  & 30  & 22  & 39  & 28  & 34  & 28  & 26  & 22  & 34  & 28  \\
\hline
4681  & H$\beta$  & 6.80 & 12.4 & 0.416 & 1.04 & 1.59 & 4.09 & 1.78 & 3.14 & 0.231 & 0.399 & 0.157 & 0.285 \\
\hline
\end{tabular}
\label{fluxes}
\end{table*}

\begin{table*}
\begin{tabular}{cccccccccccc}
\hline
$\lambda$($\AA$) & ID & \multicolumn{2}{c}{50a}  & \multicolumn{2}{c}{62} &\multicolumn{2}{c}{69} & \multicolumn{2}{c}{72} &\multicolumn{2}{c}{73a} \\
 
&    & $F_{\lambda}$ & $I_{\lambda}$ & $F_{\lambda}$ & $I_{\lambda}$ & $F_{\lambda}$ & $I_{\lambda}$ & $F_{\lambda}$ & $I_{\lambda}$  & $F_{\lambda}$ & $I_{\lambda}$ \\     
\hline
3727  & [O\,{\sc ii}] & 203 & 247 & --  & --  & 130 & 147 & 37  & 41  & --  & --  \\
4343  & H$\gamma$     & 5   & 6   & 43  & 44  & 40  & 42  & 35  & 37  & --  & --  \\ 
4861  & H$\beta$      & 100 & 100 & 100 & 100 & 100 & 100 & 100 & 100 & 100 & 100 \\
4959  & [O\,{\sc iii}]& --  & --  & 114 & 114 & 140 & 139 & 209 & 207 & 113 & 112 \\
5007  & [O\,{\sc iii}]& 60  & 58  & 334 & 332 & 417 & 410 & 608 & 599 & 314 & 310 \\ 
6563  & H$\alpha$     & 379 & 292 & 310 & 287 & 337 & 289 & 330 & 288 & 327 & 289 \\
6583  & [N\,{\sc ii}] & 94  & 72  & 43  & 40  & 28  & 24  &  -- & --  & 21  & 19  \\
6716  & [S\,{\sc ii}] & 63  & 47  & 41  & 38  & 25  & 21  &  -- & --  & 14  & 12  \\
6731  & [S\,{\sc ii}] & 60  & 45  & 22  & 20  & 17  & 14  & 43  & 37  & 11  & 9   \\
\hline
4681  & H$\beta$  & 0.0316 & 0.0717 & 0.279 & 0.354 & 0.975 & 1.58 & 0.0776 & 0.118 & 1.35 & 1.98 \\
\hline

\end{tabular}
\label{fluxes}
\end{table*}


\setlength{\bibsep}{0pt}

\bibliographystyle{mn2e}
\bibliography{ngc7793_reviewed}

\label{lastpage}

\end{document}